 \documentclass[final,5p,times,twocolumn,authornum]{elsarticle}

\usepackage{amssymb}
\usepackage{lipsum}
\usepackage{amsmath}
\usepackage{amsthm}
\usepackage{bm}
\usepackage{longtable}
\usepackage{natbib}
\usepackage{textcase}
\usepackage{url}
\usepackage{relsize}
\usepackage{amsfonts}
\usepackage{amsmath}
\usepackage{amssymb,epsf}
\usepackage{latexsym}
\usepackage{graphicx,epsfig}
\usepackage{float}
\usepackage{subfigure}
\usepackage{epstopdf}
\usepackage[colorlinks=true,citecolor=blue,linkcolor=blue,urlcolor=black]{hyperref}
\usepackage{psfrag}
\usepackage{wrapfig}
\usepackage{makeidx}
\usepackage{epsf}
\usepackage{xcolor}
\usepackage{mathtools}
\usepackage{soul}

\journal{Physics Letters B}

\begin{document}

\begin{frontmatter}



\title{Baryon asymmetry from higher-order matter contributions in gravity}


\author[first,second]{David S. Pereira}\ead{djpereira@fc.ul.pt}
\author[first,second]{Francisco S.N. Lobo}
\author[first,second]{José Mimoso}
\affiliation[first]{organization={Departamento de Física, Faculdade de Ciências da Universidade de Lisboa},
            addressline={Edifício C8}, 
            city={Campo Grande},
            postcode={1749-016 Lisboa}, 
            country={Portugal}}
\affiliation[second]{organization={Instituto de Astrofísica e Ciências do Espaço},
            addressline={Edifício C8}, 
            city={Campo Grande},
            postcode={1749-016 Lisboa}, 
            country={Portugal}}

\begin{abstract}
We investigate the observed asymmetry between matter and antimatter by incorporating higher-order matter contributions in gravity, specifically analyzing gravitational baryogenesis within the framework of $ f(R,\mathcal{T}^2) $ gravity, where $ R $ is the Ricci scalar and $ \mathcal{T}^2 \equiv T_{\mu\nu}T^{\mu\nu} $. We further explore the impact of high-order matter contributions by considering an interaction term analogous to those in gravitational and spontaneous baryogenesis, constructed using $ \mathcal{T}^2 $. The cutoff energy scale of the new interaction term is presented and its implications to Big Bang Nucleosynthesis (BBN) are discussed. The properties and implications of this term are analyzed within the frameworks of General Relativity and $f(R, \mathcal{T}^2)$ gravity. Furthermore, a connection to Big Bang Nucleosynthesis (BBN) is established, providing an observational constraint on the functional form of $f(R, \mathcal{T}^2)$.
By introducing $\mathcal{T}^2$ into the gravitational action, we propose that these modifications could significantly influence the early Universe's dynamics, thereby altering the conditions necessary for baryogenesis to occur. 
\end{abstract}


\end{frontmatter}




\section{Introduction}\label{sec:introduction}

The observed asymmetry between matter and antimatter in the primordial universe, as indicated by cosmological observations~\cite{c75ffd80-cea5-30a0-aee9-19091a4f0a9f, Burles:2000ju, WMAP:2003ivt, Burles:2000zk}, remains a fundamental open question in modern physics. This asymmetry is quantified by the baryon-to-entropy ratio, $\eta_s = n_b/s$, where $n_b$ is the net baryon number density, defined as the difference between baryon ($n_B$) and anti-baryon ($n_{\bar{B}}$) number densities, and $s$ is the radiation entropy density. The measurements from the Cosmic Microwave Background anisotropies and Big Bang Nucleosynthesis constrain this ratio to $ {n_b}/{s} = (8.8 \pm 0.6) \times 10^{-11} $~\cite{WMAP:2003ogi, Planck:2018vyg, Fields:2019pfx, ParticleDataGroup:2020ssz}.

Numerous approaches based in both particle and gravitational physics have been proposed to address this problem~\cite{Riotto:1999yt,Shaposhnikov:2009zzb,Morrissey:2012db,Pereira:2023xiw}. Among the proposed mechanisms, gravitational baryogenesis~\cite{Davoudiasl:2004gf} is particularly notable due to its dependence on gravitational interactions.
Its key feature is the interaction term~\cite{Davoudiasl:2004gf}
\begin{equation}\label{eq:GB asymmetry}
\mathcal{L}_\text{int}= \frac{\epsilon}{M_\ast^2} \left( \partial_\mu R J^\mu \right) \,,
\end{equation}
where $R$ is the spacetime Ricci curvature scalar, and $\epsilon = \pm 1$ is introduced to account for models with $ \dot{R} < 0 $ as well as $ \dot{R} > 0 $ providing the flexibility to select the appropriate sign for $ n_b $. The $M_\ast$ parameter is the cutoff energy scale typically being the reduce Plank mass $M_{Pl}\simeq2.4\times 10^{18}$ GeV, and $J^\mu$ represents normally the baryonic current, but it can represent any current that successfully leads to a net $\text{Baryon}-\text{Lepton}$ ($B-L$) charge in equilibrium, so that the asymmetry will not be wiped out by the electroweak anomaly~\cite{Davoudiasl:2004gf}.
This term, which arises from higher-dimensional operators in supergravity or quantum gravity frameworks, operates within thermal equilibrium and induces a dynamical violation of CPT symmetry.  For existing constraints on CPT violation, see~\cite{Carroll:2005dj,Li:2006ss,Li:2008tma,Xia:2008si,Mavromatos:2013vqa,Mavromatos:2013boa,McDonald:2014yfg,Mavromatos:2017gyn,Zhai:2020vob}. In the context of an expanding universe, it achieves this through a coupling between the derivative of the Ricci scalar and the baryon current $J^\mu$, resulting in a baryon asymmetry proportional to the time derivative of the Ricci scalar.

Due to its dependence on the description of gravity, gravitational baryogenesis has been studied within modified gravity frameworks,
addressing challenges such as generating baryon asymmetry during the radiation-dominated epoch, and mitigating instabilities that may arise from the implementation of the interaction term~\cite{Li:2004hh,Lambiase:2006dq,Lambiase:2012tn,MohseniSadjadi:2007qk,Bhattacharjee:2020jfk,Mojahed:2024yus,Jaybhaye:2023lgr,Baffou:2018hpe,Nozari:2018ift,Sahoo:2019pat,Odintsov:2016hgc,Arbuzova:2023rri}. These studies typically
extend the standard Einstein-Hilbert action by either introducing new geometric fields, or higher-order terms of the Ricci scalar, or even a combination of both, thereby broadening the scope of the gravitational baryogenesis mechanism. In the context of the Primordial Universe, {these} modifications to gravity are well-motivated~\cite{Nojiri:2010wj,Nojiri:2017ncd,CANTATA:2021asi}, as  at high-energy  {scales, approaching the Planck epoch}, gravity is in all likelyhood expected to deviate from General Relativity (GR), necessitating correction terms. These extensions have yielded promising scenarios such as Starobinsky inflation~\cite{Planck:2018jri}. 
  
Non-linear contributions to the fundamental action of the gravitational theory such as those involving higher-order curvature terms, that can be motivated by quantum corrections, suggest that higher-order matter contributions may also emerge in the action. Interestingly, it has been advocated that in quantum gravity if the metric can be separated into classical and quantum contributions, quantum gravity can approximate a modified gravity framework with a nonminimal interaction between gravity and matter~\cite{Dzhunushaliev:2013nea}. These findings suggest that in the context of quantum gravity matter may not only couple more intricately with geometry, but could also contribute through higher-order effects ~\cite{Maartens:2010ar,Hack:2012qf,Ford:1997hb}. 

Thus, modified theories of gravity that extend the framework of GR by incorporating higher-order matter contributions may provide critical insights into the underlying components of the Universe. In particular, during the high-energy epoch when baryogenesis is believed to occur, an extended gravitational dynamics can play a key role in the complexities of this process, potentially bridging the gap between particle physics and gravity. 
With this in mind, the present work aims {at} investigating the asymmetry between matter and antimatter through a new interaction term  
{to convey} gravitational baryogenesis, built by using the high-order matter scalar $\mathcal{T}^2 \equiv T_{\mu\nu}T^{\mu\nu}$.  The effective cutoff scale of this term is analyzed to determine its validity range and the cosmic epoch in which it may have a significant impact. The novel term will be considered both within the framework of GR and $f(R,\mathcal{T}^2)$ gravity~\cite{Katirci:2013okf}, as this theory already encompasses the $\mathcal{T}^2$ scalar.

We will consider $ f(R, \mathcal{T}^2) = R + \eta' (\mathcal{T}^2)^n $, where $\eta' \equiv \eta M_{Pl}^{2-8n}$, exploring the cases $ n = 1/2 $~\cite{Katirci:2013okf,Board:2017ign} and $ n = 1 $~\cite{Roshan:2016mbt,Board:2017ign}. 
The selection of these {$f(R,\mathcal{T}^2)$} models is motivated not only by the extensive attention they have received in the literature, where numerous constraints have been established~\cite{Katirci:2013okf,Roshan:2016mbt,Akarsu:2022abd,Akarsu:2023agp,Nazari:2022xhv,Akarsu:2018zxl,HosseiniMansoori:2023zop,HosseiniMansoori:2023mqh,Jang:2024jso}, but also by their unique cosmological properties, particularly in the context of the primordial Universe~\cite{Board:2017ign} as will be shown further below. 
The $ f(R,\mathcal{T}^2) $ theory of gravity was initially proposed as a potential explanation for the accelerated expansion of the Universe. However, beyond addressing this late-time acceleration, the theory has also been proposed to tackle other challenges such as resolving singularities in GR ~\cite{Roshan:2016mbt,HosseiniMansoori:2023zop,Barbar:2019rfn,Faraji:2021laz,Nazari:2020gnu}, and underlying Big Bang Nucleosynthesis (BBN)~\cite{Katirci:2013okf,Jang:2024jso}. 
In the specific context of baryogenesis, we argue {here} that modifications that {incorporate} non-linear matter {contributions} exhibit distinct features that can be valuable. 

The non-minimal couplings between matter and geometry can lead to gravitationally-induced particle production due to the non-conservation of the energy-momentum tensor, as discussed in~\cite{Harko:2014pqa,Harko:2015pma,Pinto:2022tlu,Pinto:2023phl,Cipriano:2023yhv}, opening possible paths to new gravity-mediated baryogenesis mechanisms, and forming a compelling basis for investigating the direct coupling of matter-related quantities with the baryonic current. However, in this work, we do not adopt gravitational particle production as the primary mechanism for generating the baryon asymmetry. Although this process can, in principle, be combined with gravitational baryogenesis, by leveraging the non-conservation of the energy-momentum tensor in non-minimally coupled theories, it demands careful treatment. 

In particular, we shall focus on gravitational baryogenesis, in which gravity does not directly cause $ B $ (or $ B-L $) violation, but rather influences as a catalist the net baryon current through the cosmological background, while the asymmetry itself arises from an independent $ B $-violating interaction. The coupling term in Equation \eqref{eq:GB asymmetry} amplifies the asymmetry to its observed value before the $ B $-violating processes decouple. 

The results presented in this work demonstrate that incorporating nonlinear matter contributions into the gravitational baryogenesis mechanism improves its effectiveness across various scenarios, including GR and modified gravity models with $n=1/2$ and $n=1$. The consistent positive outcomes support the notion that higher-order matter terms can play a significant role during high-energy regimes, particularly in the context of baryogenesis. From the cosmological effects induced by $\mathcal{T}^2$ to their impact on the generation of asymmetry, these contributions reveal noteworthy features that enhance our understanding of gravitational dynamics in the early Universe.

This paper is organized in the following manner: In Sec.~\ref{sec:T2 gravity}, we present the theoretical framework of $f(R, \mathcal{T}^2)$ gravity and the cosmological implications of the theory. In Sec.~\ref{sec:GB T2}, we provide the theoretical aspects of gravitational baryogenesis and its extensions in the context of $f(R, \mathcal{T}^2)$ gravity applying such results for the models under consideration. Finally, Sec.~\ref{sec:Summary and conclusion} provides a summary and discussion of our results.

\section{Baryogenesis with $\mathcal{T}^2$}\label{sec:GB T2}

As mentioned earlier, we will consider a new interaction term, {{analogous to}} the original one \eqref{eq:GB asymmetry}, built using the scalar $\mathcal{T}^2$. This term is given by
\begin{equation}\label{eq:T2 asymmetry}
    \mathcal{L}_{\text{int}} = \frac{\epsilon}{M_{\ast}^8} \partial_\mu \left(\mathcal{T}^2\right)J_B^{\mu} \,,
\end{equation}
that leads to the asymmetry
\begin{equation}\label{eq:asymmetry nb T2}
    \frac{n_b}{s} \simeq - \frac{15g_b\epsilon}{4\pi^2 g_\ast} \frac{\dot{\mathcal{T}}^2}{M^8_\ast T} \Bigg|_{T_D}\,,
\end{equation}
where $ g_{\ast} $ denotes the total effective number of relativistic degrees of freedom~\cite{Kolb:1990vq} and $g_b$ is the number of internal degrees of freedom of baryons and for this work, we will consider $g_b \sim \mathcal{O}(1)$, $g_\ast=106.75$. 

This interaction term \eqref{eq:T2 asymmetry} may arise in non-minimal couplings between matter and gravity, particularly within effective actions for quantum field theory in curved spacetime. Additionally, it can be motivated by braneworld scenarios~\cite{Maartens:2010ar} and potentially by string-inspired effective actions~\cite{Green:1987sp,Green:1987mn}. More specifically, in the low-energy effective action for D-branes, this term could emerge as a consequence of the backreaction of matter fields on the brane geometry. Furthermore, considering a cosmological context where the energy-momentum tensor is describe by a perfect fluid
\begin{equation}\label{eq:Perfect fluid}
    T_{\mu\nu} = (\rho + p) u_\mu u_\nu + pg_{\mu\nu} \, ,
\end{equation}
where $\rho$ is the energy density and $p$ the isotropic pressure, the scalar $\mathcal{T}^2$, considering $p=w\rho$, where $w$ is constant, is given by 
\begin{equation}\label{eq:T2}
    \mathcal{T}^2 = \rho^2\left(1 + 3w^2\right) \,,
\end{equation}
allowing to rewrite Eq.\eqref{eq:T2 asymmetry} as
\begin{equation}\label{eq: T2 rho2}
    \mathcal{L}_{\text{int}} = \frac{\epsilon\left(1+3w^2\right)}{M_{\ast}^8} \partial_\mu \left(\rho^2\right)J_B^{\mu} \,,
\end{equation}
with the dependence on $\rho^2$ clearly demonstrating a higher-order contribution from the matter sector. From a theoretical standpoint, the presence of the $\rho^2$ term naturally arises in cosmological equations derived from various frameworks, including braneworld scenarios~\cite{Shtanov:2002mb} and quantum loop gravity~\cite{Ashtekar:2006uz}, hence, it is reasonable to argue that the term \eqref{eq:T2 asymmetry} may also appear in these frameworks, particularly in the high-energy regime of the Primordial Universe, where quantum gravitational effects become significant. 

A fundamental point to consider when working with beyond the standard physics in the early Universe is its impact on BBN. To assess this, we compute the overall cutoff scale of the dimension-12 Effective Field Theory (EFT) operator given in Eq.~\eqref{eq:T2 asymmetry}, obtaining
\begin{equation}
    \Lambda = M_{\ast}\,,
\end{equation}
hence, to maintain the validity of the interaction term, we will impose $T_D< M_\ast$. Furthermore, if the relevant energy scale of the process is $E$, dimensional consistency suggests the interaction contributes as
\begin{equation}
\mathcal{L}_{\text{int}} \sim \epsilon \left(\frac{E}{M_\ast}\right)^8 E^4,
\end{equation}
since the Lagrangian density has mass dimension 4. As this term is expected to act in the high energy regimes, $M_\ast$ is expected to {{be}} large, therefore, significant effects from this term arise only at correspondingly high temperatures. Notably, in what concerns BBN, for the range of temperatures possible for BBN, $T_{BBN}\sim 1 $MeV, even a relatively low cutoff of $M_\ast \sim 10^{7}$ GeV leads to
\begin{equation}
\left(\frac{T_{\text{BBN}}}{M_\ast}\right)^8 = \left(\frac{10^{-3}\,\text{GeV}}{10^{7}\,\text{GeV}}\right)^8 = 10^{-80}\,,
\end{equation}
translating in a massive suppression, rendering the contribution from this operator negligible at BBN energies. Although this term is suppressed during the BBN epoch, it may still have an impact on primordial gravitational waves, both independently and through couplings with other baryogenesis mechanisms. This phenomenon will be studied in detail in future work. Furthermore, one may also contemplate the potential back-reaction effects of the effective Lagrangian \eqref{eq:T2 asymmetry}, which could induce noteworthy deviations from standard cosmology. However, a detailed investigation of these effects lies beyond the scope of this Letter.

With this settled, analyzing now Eq.\;\eqref{eq: T2 rho2} we can extract important conclusions. Considering that baryogenesis occurs during the radiation epoch, the dependence on $(1+3w^2)$ of Eq.\;\eqref{eq: T2 rho2}, which remains nonzero during the radiation-dominated epoch ($w=1/3$), stands in contrast to the dependence on $(1-3w)$~\cite{Davoudiasl:2004gf}, which characterizes \eqref{eq:GB asymmetry}, and vanishes in the radiation epoch. Furthermore, even when accounting for interactions among massless particles, where typical gauge groups and matter content at high energies yield values of $1 - 3w $ in the range $10^{-2} - 10^{-1}$~\cite{Davoudiasl:2004gf}, and taking the maximum decoupling temperature as the upper bound on the inflationary energy scale, approximately $M_I \simeq 1.6 \times 10^{16}$ GeV, based on constraints from tensor mode fluctuations~\cite{Planck:2018jri}, the original gravitational baryogenesis scenario within General Relativity remains unsuccessful. This unsuccess also occurs for a more realistic scenario, where one considers $T_D\le T_{RD} < M_I$ with $T_{RD}$ being the temperature at which the universe becomes radiation dominated, i.e., the reheat temperature in this case. For the new interaction term, due to the dependence on $(1+3w^2)$ and the quadratic energy density term, these problems are expected to be solved.

To better comprehend the impact on baryogenesis of the new interaction term, we will study it in the context of GR and $f(R,\mathcal{T}^2)$ gravity, as this modified theory of gravity uses the $\mathcal{T}^2$ scalar into the gravitational description having a possible link to the origin of Eq.\eqref{eq:T2 asymmetry}.

\section{$f(R,\mathcal{T}^2)$ gravity}\label{sec:T2 gravity}
\subsection{Action and Field Equations}
The action for $f(R,\mathcal{T}^2)$  gravity reads
\begin{equation}\label{eq:actionsf(R,T2)}
    S=\frac{1}{2\kappa}\int d^4x \sqrt{-g} f(R,\mathcal{T}^2) +\int d^4x \sqrt{-g} \mathcal{L}_m \, ,
\end{equation}
where $\kappa = 8\pi G = M_{Pl}^{-2}$ and $f(R,\mathcal{T}^2)$ is a well-behaved function of the Ricci scalar and $\mathcal{T}^2$.

We define the energy-momentum tensor of the matter fields as 
\begin{equation}\label{Stresse-energy tensor}
    T_{\mu\nu} = -\frac{2}{\sqrt{-g}}\frac{\delta (\sqrt{-g}\mathcal{L}_M)}{\delta g^{\mu\nu}}\, , 
\end{equation}
and imposing that $\mathcal{L}_m$ depends solely on the metric components and not on their derivatives we obtain
\begin{equation}
     T_{\mu\nu} = g_{\mu\nu} \mathcal{L}_m - \frac{\partial \mathcal{L}_m }{\partial g^{\mu\nu} } \, .
\end{equation}

Varying the action \eqref{eq:actionsf(R,T2)} with respect to the metric yields the field equations 
\begin{eqnarray}\label{FE actionsf(R,T2)}
	f_{,R} R_{\mu\nu} -\frac{1}{2}g_{\mu\nu}f +  (g_{\mu\nu} \nabla^\alpha \nabla_\alpha -\nabla_\mu \nabla_\nu)f_{,R} \nonumber \\
 = \kappa \left(T_{\mu\nu} - \frac{1}{\kappa}f_{\mathcal{T}^2} \theta_{\mu\nu}\right) \, ,
\end{eqnarray}
where subscripts denote differentiation with respect to the associated quantity and $\theta_{\mu\nu}$ is defined as
\begin{eqnarray}
	\theta_{\mu\nu} \equiv \frac{\delta T^{\alpha\beta} T_{\alpha\beta}}{\delta g^{\mu\nu}} 
	 = 
	-2\mathcal{L}_m \left(T_{\mu\nu} - \frac{1}{2}g_{\mu\nu}T \right) - TT_{\mu\nu}  
	\nonumber  \\ 
	 + 2T^{\alpha}_\mu T_{\nu\alpha} 
    - 4T^{\alpha\beta} \frac{\partial^2 \mathcal{L}_m}{\partial g^{\mu\nu} \partial g^{\alpha\beta}} ,
\end{eqnarray}
where $ T $ is the trace of the standard energy-momentum tensor.

The trace of the field equations then becomes  
\begin{equation}\label{eq:FE trace}
2 \left( \mathcal{T}^2 \right)^n - n \left( \mathcal{T}^2 \right)^{n-1} \theta = -\frac{1}{\eta'} \left( \kappa T + R \right) \,,
\end{equation}
where $ \theta $ is the trace of $ \theta_{\mu\nu} $. This equation allows the evaluation of $ \mathcal{T}^2 $ based on the choice of the parameter $ n $.

In this work, we adopt the choice $\mathcal{L}_m = p$~\cite{Katirci:2013okf, Bertolami:2007gv, Schutz:1970my, Bertolami:2008ab}, as it provides a thermodynamically accurate description of the early Universe during the baryogenesis epoch. This approach effectively captures the energy properties of the primordial Universe and allows for detailed analysis of gravitational baryogenesis. To mitigate divergences caused by $ \frac{\partial^2 \mathcal{L}_m}{\partial g^{\mu\nu} \partial g^{\alpha\beta}} $ due to $p=0$, we set this term to zero, consistent with existing literature (see~\cite{Akarsu:2023lre} for more in-depth analysis of this choice). Consequently, the tensor $ \theta_{\mu\nu} $ is given by~\cite{Akarsu:2023lre}  
\begin{equation}
\theta_{\mu\nu} = -2\mathcal{L}_m \left(T_{\mu\nu} - \frac{1}{2}g_{\mu\nu}T\right) - T T_{\mu\nu} + 2T^\alpha_\mu T_{\nu\alpha} \,.
\end{equation}

With this clarified, for the matter sector, we assume that it is described by a perfect fluid~\eqref{eq:Perfect fluid} allowing to write the trace of the energy-momentum tensor  and the tensor $\theta_{\mu\nu}$ as
\begin{equation}\label{eq:T}
    T= (3w-1)\rho \, ,
\end{equation}
\begin{equation}
     \theta_{\mu\nu}=-(\rho^2 + 4p\rho + 3p^2) u_\mu u_\nu,
\end{equation}
respectively.

\subsection{Cosmology}

We will consider that the asymmetry generation occurs after the inflationary period to avoid the washout problem caused by the rapid expansion (for discussions on scenarios where baryogenesis is considered to occur during inflation or immediately after its end, see~\cite{Pereira:2024ddu, Ahmad:2019jbm, Rangarajan:2001yu,Cado:2023zbm,Gunji:2023xme}). Hence, we consider a flat Friedmann-Lemaître-Robertson-Walker (FLRW) metric 
\begin{equation}\label{eq:FLRW metric}
 \textrm{d}s^{2}=- \textrm{d}t^{2}+a^{2}(t)\textrm{d}V^2 ,  
\end{equation}
where $t$ is the comoving proper time, $a(t)$ is the expansion scale factor and $\textrm{d}V$ is the volume element in comoving coordinates.

Using this metric, the modified Friedmann equations can be expressed as~\cite{Board:2017ign}:  
\begin{equation}\label{friedmann}
H^2 = \kappa \frac{\rho}{3} + \frac{\eta'}{3} (\rho^2 + 3p^2)^{n-1} \left[(n - \frac{1}{2})(\rho^2 + 3p^2) + 4n\rho p \right] \, ,
\end{equation}
where $ H \equiv \dot{a}/a $ is the Hubble parameter. The corresponding acceleration equation is given by:
\begin{eqnarray}\label{acceleration}
\dot{H} + H^2 = - \kappa \, \frac{\rho + 3p}{6} - \frac{\eta'}{3} (\rho^2 + 3p^2)^{n-1} \times \nonumber \\ 
\times \left[\frac{n+1}{2}(\rho^2 + 3p^2) + 2n\rho p \right] \, .
\end{eqnarray}

Assuming that the matter fields satisfy a barotropic equation of state $ p = w\rho $, where $ w $ is a constant, the additional terms introduced by the scalar $ \mathcal{T}^2 $ are proportional to $ \rho^{2n} $, each scaled by a constant. Consequently, the cosmological equations can be reformulated as~\cite{Board:2017ign}
\begin{equation}\label{eq:Fridmann T^2}
H^2 = \kappa \frac{\rho}{3} + \frac{\eta' \rho^{2n}}{3} F_{\text{Frd}}(n, w) \, ,
\end{equation}
where $ F_{\text{Frd}} $ depends on $ n $ and $ w $, and is explicitly defined as:  
\begin{equation}\label{Acc T^2}
F_{\text{Frd}}(n, w) \equiv \left(1 + 3w^2\right)^{n-1} \left[\left(n - \frac{1}{2}\right)\left(1 + 3w^2\right) + 4nw \right] \, .
\end{equation}

Similarly, the acceleration equation becomes:  
\begin{equation}\label{eq:Acel}
\dot{H} + H^2 = -\kappa \frac{1 + 3w}{6}\rho - \frac{\eta' \rho^{2n}}{3} F_{\text{Acc}}(n, w) \, ,  
\end{equation}
where $ F_{\text{Acc}} $, which also depends on $ n $ and $ w $, is given by:
\begin{equation}
F_{\text{Acc}}(n, w) \equiv \left(1 + 3w^2\right)^{n-1} \left[\frac{n+1}{2}\left(1 + 3w^2\right) + 2nw \right] \, .
\end{equation}

The modified continuity equation is derived from the Friedmann equation, leading to  
\begin{equation}\label{eq:modified continuity}
    \dot{\rho} = -3H\rho(1+w)F_{con}(n,w) \, ,  
\end{equation}  
where $ F_{con}(n,w) $ represents an additional term arising from $ \mathcal{T}^2 $ and is defined as  
\begin{equation}\label{eq:const continuity}
    F_{con}(n,w) \equiv \frac{\kappa+\eta'\rho^{2n-1} n(1+3w)(1+3w^2)^{n-1}}{\kappa+2\eta'\rho^{2n-1}n F_{Frd}(n,w)} \, .  
\end{equation}  

An interesting feature of this theory is the term $ \rho^{2n} $, present in Eq. \eqref{eq:Fridmann T^2} and Eq. \eqref{eq:Acel}, that is reminiscent of quantum geometric effects observed in loop quantum gravity~\cite{Ashtekar:2006uz} and in braneworld scenarios~\cite{Shtanov:2002mb}.

We will consider that baryogenesis occurs during the radiation-dominated epoch, relevant for baryogenesis, $ w = 1/3 $. However, this choice may not yield the standard energy density form, $ \rho \propto a^{-4} $, due to the modification of the continuity equation~(\ref{eq:modified continuity},\ref{eq:const continuity}). Therefore, to address this point, one can impose
\begin{equation}
    (1+w)F_{con}(n,w) = \frac{4}{3} \, ,  
\end{equation}  
which constrains $ \rho^{2n-1} $ as
\begin{equation}\label{eq:rho constrangido}
    \rho^{2n-1} = \frac{1-3w}{\eta M_{\text{Pl}}^{4-8n} \left( 3nA(w) - 8nF_{Frd}\right)} \, ,  
\end{equation}  
with $ A(w) \equiv (1+w)(1+3w)(1+3w^2)^{n-1} $. For a physical and dynamical cosmological evolution, $ n=1/2 $ is required, ensuring $ \rho \propto a^{-4} $ for specific $ \eta $ and $ w $. This leads to the relation  
\begin{equation}\label{eq:eta constrangido}
    \eta_\ast \equiv \eta(w) = \frac{1-3w}{\frac{3}{2}A(w) - 4F_{Frd}(1/2,w)} \, ,  
\end{equation}  
which equates $ \eta $ based on $ w $. This relation not only ensures the desired energy density form, but also imposes a constraint on $ \eta $. The value of $ w $ determines the corresponding $ \eta $, opening new avenues for studying baryogenesis. Notably, for $ w = 1/3 $, $ \eta $ vanishes, reducing the theory to GR, a result that will be analyzed in detail in the next section. 

In this study, we will only consider the case $ w=1 $, which corresponds to stiff matter, exploring its impact on baryogenesis. In standard cosmology, a stiff fluid with $ w=1 $ dilutes very quickly as $ \rho \propto a^{-6} $. However, in our case, the energy density behaves like radiation, scaling as $ \rho \propto a^{-4} $, which means the stiff component remains relevant for a longer period. This can have interesting consequences, such as affecting the reheating process, enhancing the gravitational wave background, and potentially influencing late-time cosmology~\cite{Gouttenoire:2021jhk}.

In the following analysis, $ w=1/3 $ will be fixed for all models, while for $ n=1/2 $, $ w=1$ will be also explored via Eq.~\eqref{eq:eta constrangido}. The model's choice, particularly the power of $ \mathcal{T}^2 $, plays a crucial role in shaping the cosmological framework and its parameters, with $ \eta $ being central to the gravitational baryogenesis mechanism. With this being said, in order to compute the asymmetries associated with Eq. \eqref{eq:GB asymmetry} and Eq. \eqref{eq:T2 asymmetry}, it is necessary to determine the expressions for $ H(t) $ and $ \rho $, both of which depend on the cosmological equations specific to each model ($ n = \frac{1}{2} $ and $ n = 1 $). Accordingly, we will use Eqs. \eqref{eq:Fridmann T^2}, \eqref{eq:Acel}, and \eqref{eq:modified continuity} to derive their analytical forms.

\section{Effects of high-order matter contributions on baryogenesis}

In this section, we conduct a detailed analysis of the impact of higher-order matter contributions on baryogenesis. Prior to this, is is important to establish a connection between baryogenesis and BBN to assess whether the proposed baryogenesis mechanisms can generate the observed baryon asymmetry without disrupting BBN. This connection further provides a means to constrain the $f(R,\mathcal{T}^2)$ gravity theory. 

BBN occurs at temperatures around $ 0.1 - 100 $ MeV, far after baryogenesis, following the freeze-out of weak interactions near $ T_\text{freeze} \approx 0.6$ MeV in standard cosmology~\cite{Yeh:2022heq,Cyburt:2015mya,Park:2022lgf,Mukhanov:2003xs,Capozziello:2017bxm,Asimakis:2021yct}. However, in our framework, modifications to the Hubble parameter create an influence in the timing of element formation. The initial conditions necessary for nucleosynthesis are established within a relativistic plasma, where weak interactions govern the neutron-proton conversion processes: 
\begin{equation}
\label{reactions}
\begin{split}
& e^+ + n \longleftrightarrow p + \bar{\nu_e} ~,\\
& \nu_e + n \longleftrightarrow p + e^-~,\\
& n \longleftrightarrow p + e^- + \bar{\nu_e}~.
\end{split}
\end{equation}

The weak interaction rate, $\Lambda(T)$, which determines the efficiency of these conversion processes, is then given by (see ~\cite{Capozziello:2017bxm,Asimakis:2021yct,Ghoshal:2021ief} for details)
\begin{equation}
\Lambda(T) = 4AT^4\left(4!T^2+2\times3!QT+2!Q^2\right) \simeq c_qT^5  + \mathcal{O}\left(\frac{Q}{T}\right),
\end{equation}  
where $Q=m_n-m_p$ is the mass difference of neutron and proton, $A=1.02\times10^{-11}\text{GeV}^{-4}$ and $c_q=4A4!\simeq 9.8\times^{-10}\text{GeV}^{-4}$. The freeze-out temperature, $ T_{\text{freeze}} $, is therefore determined by the condition  
\begin{equation}\label{eq:FreezCondition}
 \Lambda(T_\text{freeze}) \simeq H(T_\text{freeze})\,,
\end{equation}  
where the interaction rate equals the cosmic expansion rate and in General Relativity we have
\begin{equation}
H_{\text{GR}}=\sqrt{\frac{\rho_{\text{radiation}}}{3M_{Pl}^2}}\,,
\end{equation}
where the energy density of relativistic species is given by~\cite{Kolb:1990vq}
\begin{equation}\label{eq:rho vs temperature}
    \rho_{\text{radiation}}=\frac{\pi^2}{30}g_\ast T^4 \, ,
\end{equation}
giving the decoupling temperature $T_{\text{freeze}}\approx 0.6 \ \text{MeV}$~\cite{Capozziello:2017bxm,Asimakis:2021yct,Ghoshal:2021ief}. This defines the onset of nucleosynthesis as weak interactions become inefficient. Modifications to gravity alter the expansion rate, which impacts the freezing temperature and consequently the primordial helium mass fraction, $ Y_p $, as it is sensible to $T_{\text{freeze}}$. The deviation $ \delta Y_p $ is constrained to be below $ 10^{-4} $~\cite{Coc:2003ce,Izotov:2003xn}, imposing an upper bound on the relative variation of the freeze-out temperature~\cite{Capozziello:2017bxm,Asimakis:2021yct} 
\begin{equation}
\left|\frac{\delta T_{\text{freeze}}}{T_{\text{freeze}}}\right|<4.7\times 10^{-4}\,.
\end{equation}

The effects of the scalar $ \mathcal{T}^2 $ on the expansion rate can be analyzed at the eyes of this constraint by considering that in the context of modified theories of gravity one has
\begin{equation}\label{eq: H deviations}
    H = H_{\text{GR}} + \delta H\,,
\end{equation}
where $\delta H$ is the variation in the standard expansion introduced by the modifications in question. Using Eq.\eqref{eq:FreezCondition} one derives~\cite{Capozziello:2017bxm}
\begin{equation}
    \delta H = 5c_q T^4_{\text{freeze}} \delta T_{\text{freeze}}\,,
\end{equation}
leading to the relation
\begin{equation}\label{eq:T vs dH}
    \frac{\delta T_{\text{freeze}}}{T_{\text{freeze}}} = \frac{\delta H}{5c_q T^5_{\text{freeze}}}\,.
\end{equation}

In the following subsections, we will apply the previous equation to evaluate whether the obtained baryogenesis scenarios satisfy BBN constraints, thereby also constraining the $f(R, \mathcal{T}^2)$ models under consideration.
\subsection{Coupling between $\partial_\mu (R)$ and $J^\mu$}\label{subsec:Rcoup}

{\it (i) Model $n=1/2$}:\label{subsec:R n 1/2}

For the case with $n=1/2$ the modified cosmological equations read
\begin{equation}\label{eq:Friedmann n=1/2}
    H^2= \left( \frac{3+\eta \sqrt{3}}{9}\right)\frac{\rho}{M_{Pl}^{2}} \, ,
\end{equation}
\begin{equation}\label{eq:Acceleration n=1/2}
   \dot{H} + H^2= -\left(\frac{3+\eta 2\sqrt{3}}{9}\right)\frac{\rho}{M_{Pl}^{2}} \,,
\end{equation}

By substituting Eq. \eqref{eq:Friedmann n=1/2} into Eq. \eqref{eq:Acceleration n=1/2} yields the differential equation
\begin{equation}
    \dot{H} + \left( \frac{3\eta+ 2\sqrt{3}}{\eta + \sqrt{3}}\right)H^2 = 0 \, ,
\end{equation}
that has the analytical solution 
\begin{equation}\label{eq:H n=1/2}
    H(t) = \alpha(\eta) t^{-1} \, , 
\end{equation}
with $ \alpha(\eta) = \frac{\eta + \sqrt{3}}{3\eta + 2\sqrt{3}}$. The obtained Hubble parameter corresponds to a power law for the scalar factor, $a(t)\sim t^\alpha$ that depends on $\eta$ with $\alpha(\eta=0)=\frac{1}{2}$ recovering the usual expansion during the radiation epoch. Utilizing the solution derived for $H$, the time derivative of the Ricci scalar is given by
\begin{equation}\label{eq:Ricci dot n=1/2}
    \dot{R} = -12\alpha(\eta) \left(\frac{2\alpha(\eta) - 1}{t^3} \right) \, .
\end{equation}

The solution for the Hubble parameter also allows to derive a time-dependent expression for $\rho$. By substituting this solution into the continuity equation \eqref{eq:modified continuity}, we obtain 
\begin{equation}
    \dot{\rho}(t) + 2t^{-1}\rho(t) = 0 \, ,
\end{equation}
that has the solution
\begin{equation}\label{eq:rho n=1/2 R}
    \rho(t) = \rho_0 t^{-2}\, ,
\end{equation}
where $ \rho_0 $ represents a constant. Examining Eq.\eqref{eq:rho n=1/2 R} we see that it corresponds to the standard description of energy density in a radiation-dominated universe in GR with $H=\frac{1}{2}t^{-1}$. This is an intriguing characteristic of this model, as, despite introducing modifications to the continuity equation, it still allows the recovery of the usual energy density behavior when $w = 1/3$. This occurs because the deviations in the expansion rate, as given by Eq. \eqref{eq:H n=1/2}, effectively counterbalance the introduced modifications. Furthermore, deviations introduced by the $\mathcal{T}^2$ term are expected to manifest in $\rho_0$, reflecting its impact on the overall dynamics. By substituting the solution for $ \rho_0 $ alongside the solution for the Hubble parameter $ H $ into the Friedmann equation \eqref{eq:Friedmann n=1/2}, we obtain
\begin{equation}\label{eq:rho_0 R n=1-2}
    \rho_0(\eta) = \frac{3\sqrt{3}\eta + 9}{\left(3\eta + 2\sqrt{3}\right)^2} M_{\text{Pl}}^2 \, .
\end{equation}  

We can relate the time of decoupling with the temperature of decoupling by using the equation that relates the total radiation density with the energy of all relativistic species, Eq.\eqref{eq:rho vs temperature}, and solving for $t_D$ allows to write
\begin{equation}\label{eq:t_D n=1/2}
    t_D = \left(\frac{30\rho_0}{\pi^2 g_\ast}\right)^{1/2} \ T^{-2}_D \, .
\end{equation}
that in combination with Eq.\eqref{eq:Ricci dot n=1/2} and Eq.\eqref{eq:GB asymmetry} results in
\begin{equation}\label{eq:Asymmetry final R 1/2}
     \frac{n_b}{s} \simeq \epsilon\frac{45\pi  \alpha(\eta) \left(2\alpha(\eta)-1\right) g^{1/2}_s g_b }{M^2_\ast (30 \rho_0)^{3/2}} T_D^5 \, .
\end{equation}

Before conducting a numerical analysis, it is crucial to address an important theoretical remark. If an inflationary behavior is considered then, in order to avoid ghosts and gradient instabilities, one must impose $ 2 > \eta > 0 $~\cite{HosseiniMansoori:2023zop} (noting that our model is defined with $ +\eta $). This consideration can, to some extent, be disregarded. However, from a theoretical perspective and in light of constraints on the parameter $\eta$, we will address it in a general manner. If scenarios of successful baryogenesis are identified outside the allowed $\eta$ domain, they will be presented. However, we would like to stress that they are subjected to the caveat that these cases are incompatible with the model's inflationary component.
\begin{figure}[t!]
\hspace{-1.5em}
    \includegraphics[width=0.475\textwidth]{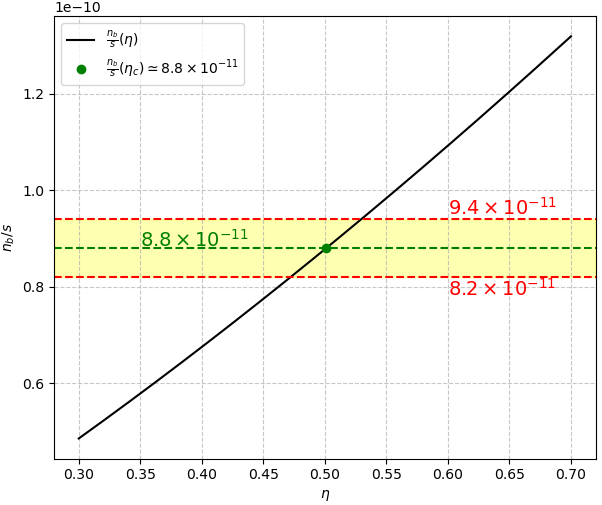} 
    \caption{Plot of $\frac{n_b}{s}$ vs $\eta$. The black line shows the evolution of $\frac{n_b}{s}(\eta)$, the dashed red lines indicate the observation uncertainties for $\frac{n_b}{s}$, with the yellow band representing a validity zone, while the green line marks the observation value for the asymmetry. The green dot marks the point where $\frac{n_b}{s}(\eta_{c}) \simeq 8.8\times 10^{-11}$ with $\eta_{c} =0.50112$.}
  \label{Figure_1}
  \end{figure}

Within the constrained interval for $\eta$, $ 2\alpha(\eta) - 1 < 0 $, we are lead to set $ \epsilon = -1 $. By numerically analyzing Eq.~\eqref{eq:Asymmetry final R 1/2}, we identify a relationship between $ \eta $ and $ T_D $, where higher temperatures correspond to lower values of $ \eta $ necessary to achieve a successful asymmetry. Specifically, for $ M_\ast = 10^{16} \, \text{GeV} $ and $ T_D = 2.5 \times 10^{15} \, \text{GeV} $, the dependence of the asymmetry on $\eta$ is illustrated in Fig.~\ref{Figure_1}. Analyzing this figure, we can see that the asymmetry grows linearly with $\eta$ demonstrating the impact of the additional terms introduced by $\mathcal{T}^2$.

Analyzing now using the BBN constraints, from Eq.\eqref{eq:Friedmann n=1/2} we obtain
\begin{equation}\label{eq: BBN 1/2}
    \delta H = \left(\sqrt{1+\frac{\eta}{\sqrt{3}}}-1\right)H_{\text{GR}} 
\end{equation}
leading to 
\begin{equation}
    \frac{\delta T_{\text{freeze}}}{T_{\text{freeze}}} = \frac{\left(\sqrt{1+\frac{\eta}{\sqrt{3}}}-1\right)\left(\pi \sqrt{g_{\text{BBN}}}\right)}{15\sqrt{10}c_q T^3_{\text{freeze}}M_{Pl}}<4.7\times 10^{-4}\,, 
\end{equation}
where $g_{BBN} \approx 10$, leading to the upper bound
\begin{equation}
\eta<0.00395156 \xrightarrow{} \eta' < 6.860347\times 10^{-40},
\end{equation}
which excludes the value $\eta_c = 0.50112$ as a viable scenario under BBN constraints. However, for $M_\ast = 10^{17}$ GeV and $T_D = 2\times 10^{16}$ GeV we found $\frac{n_b}{s}(\eta=0.00192932)=8.0003\times 10^{-11}$, a scenario in which both the observed asymmetry and the BBN constraints are fulfilled. 

Although both values of $ \eta_c $ are much smaller than $M_{Pl}$, that is, smaller than the Ricci coupling constant, the contribution from $\mathcal{T}^2$ is high enough that it disturbs BBN even with a small coupling constant. This reflects that the small contribution of the additional term $ \mathcal{T}^2 $ can significantly affect the system. The influence of $ \mathcal{T}^2 $ primarily emerges from the additional terms $ \frac{\eta'\sqrt{3}}{9} $ and $ 2\frac{\eta'\sqrt{3}}{9} $ in the corresponding cosmological equations, introducing novel dynamics into the cosmological framework. These deviations from GR are minimal for the obtained value of $ \eta $, indicating that resolving the asymmetry problem does not require substantial or complex departures from GR. Furthermore, it is possible to construct a model where $ \eta $ depends on time and evolves such that, at a specific scale, $ \eta(t) = 0 $, reducing the model to GR.

Incorporating now the constraint from Eq.\eqref{eq:eta constrangido}, the cosmological equations take the following form  
\begin{equation}\label{Friedmann n=1/2 constraint}
    H^2 = \omega(\eta_\ast, 1) \frac{\rho}{3M^2_{Pl}} \, ,  
\end{equation}  
\begin{equation}\label{Acc n=1/2 constraint}
    \dot{H} + H^2 = -\sigma(\eta_\ast, 1) \frac{\rho}{3M^2_{Pl}} \, ,  
\end{equation}  
where $ \eta_\ast $ corresponds to the value of $ \eta $ derived from Eq.\eqref{eq:eta constrangido}, and the parameters $ \omega(\eta_\ast, 1) $ and $ \sigma(\eta_\ast, 1) $ are defined as  
\begin{equation}
     \omega(\eta_\ast, 1) \equiv 1 + \eta_\ast F_{\mathrm{Frd}}\left(\frac{1}{2}, 1\right) \, ,  
\end{equation}  
\begin{equation}
    \sigma(\eta_\ast, w) \equiv \frac{3 + 2\eta_\ast F_{\mathrm{Acc}}\left(\frac{1}{2}, 1\right)}{2} \, .  
\end{equation}  

Substituting Eq.~\eqref{Friedmann n=1/2 constraint} into Eq.~\eqref{Acc n=1/2 constraint} yields the differential equation:  
\begin{equation}
    \dot{H} + \frac{\omega(\eta_\ast, 1) + \sigma(\eta_\ast, 1)}{\omega(\eta_\ast, 1)} H^2 = 0 \, ,  
\end{equation}  
whose solution has the same form of Eq.\eqref{eq:H n=1/2} with $\lambda(\eta_\ast, 1) \equiv {\omega(\eta_\ast, 1)}/\left({\omega(\eta_\ast, 1) + \sigma(\eta_\ast, 1)}\right)$ instead of $\alpha$. Additionally, $\dot{R}$ will have the same form as in Eq.\eqref{eq:Ricci dot n=1/2} now with $\lambda$. 

A key distinction in this constrained case lies in the continuity equation, now given by  
\begin{equation}
    \dot{\rho} = -4H\rho \, ,  
\end{equation}  
which has the solution  
\begin{equation}\label{rho 1-2 constrangido}
    \rho = \rho_c t^{-4\lambda} \, ,  
\end{equation}  
where $ \rho_c $ is a positive constant with units of $ \text{GeV}^{4 - 4\lambda} $.  

Using Eq.~\eqref{eq:rho vs temperature}, the decoupling time-temperature relation becomes  
\begin{equation}\label{eq:time 1-2 constraint}
    t_D = \left(\frac{\pi^2 g_\ast}{30}\right)^{-1/{4\lambda}} T_D^{-1/\lambda} \rho_c^{1/4\lambda} \, ,  
\end{equation}  
giving the asymmetry  
\begin{equation}
    \frac{n_B}{s} \simeq \epsilon\frac{45g_b (\pi^2 g_\ast)^{\frac{3}{4\lambda} - 1} \lambda(2\lambda - 1)}{30^{\frac{3}{4\lambda}} M_\ast^2 \rho_c^{\frac{3}{4\lambda}}} T_D^{\frac{3}{\lambda} - 1} \, .  
\end{equation} 

Considering the parameter values $ T_D = 1 \times 10^8 $, $ M_\ast = M_{Pl} $, $\epsilon=-1$ and $ \rho_c = 9.545 \times 10^{57} $, our analysis yields $ n_b/s \simeq 8.8005 \times 10^{-11} $. However, from the BBN constraint, using $\omega(\eta_\ast,1)$ instead of $\sqrt{1+\eta/\sqrt{3}}$ in Eq.\eqref{eq: BBN 1/2}, a possible feat due to the fact that the energy density of the scalar field behaves as the energy density of radiation because the condition \eqref{eq:eta constrangido}, we found that $\eta_\ast (1) = -0.125$ violates the bound established making this case not viable for BBN.

All the obtained results, combined with the natural emergence of this model within the standard energy density evolution for radiation-dominated epochs, emphasize its significance. Furthermore, the role of $\mathcal{T}^2$ is governed by an effective coupling constant that depends solely on $\eta$ and, given that $\eta$ is small, this underscores how subtle modifications to GR can effectively tackle the baryon asymmetry problem. This framework highlights the potential to resolve this longstanding challenge through minimal yet impactful adjustments to the standard GR framework.\\
{\it (ii) Model $n=1$}:\label{sec:R n 1}

The $ n = 1 $ case results in the following modified Friedmann equations  
\begin{equation}\label{eq:Friedmann n=1}
    H^2 = \frac{1}{3M_{Pl}^2}\rho + \frac{2}{3}\frac{\eta}{M_{Pl}^{6}} \rho^2 \, ,  
\end{equation}  
\begin{equation}\label{eq:Acc n=1}
    \dot{H} + H^2 = -\frac{1}{3M_{Pl}^2}\rho - \frac{2}{3}\frac{\eta}{M_{Pl}^{6}} \rho^2 \, ,  
\end{equation}  
respectively.  

Substituting Eq.~\eqref{eq:Friedmann n=1} into Eq.~\eqref{eq:Acc n=1} simplifies to the differential equation  
\begin{equation}
    \dot{H} + 2H^2 = 0 \, ,  
\end{equation}  
which has the solution 
\begin{equation}\label{eq:H n=1}
    H(t) = \frac{1}{2}t^{-1} \, ,  
\end{equation}  
implying a scale factor evolution $ a \propto t^{1/2} $. This Hubble parameter results in $ R = 0$ and $ \dot{R} = 0 $, indicating that no asymmetry is generated via Eq.~\eqref{eq:GB asymmetry} in this scenario. Although the Friedmann equations for this case exhibit significant deviations from GR, due to the quadratic term $ \rho^2 $, from the perspective of baryogenesis, this case behaves identically to the $ w = 1/3 $ case in GR. The expansion rate behaves in in the same way as standard GR. Although the expansion rate is the same as in GR, the exotic continuity equation will impact the overall cosmological dynamics.

Applying Eq.\eqref{eq: H deviations} into Eq.\eqref{eq:Friedmann n=1} one obtains
\begin{equation}
    \delta H = \left(\sqrt{1 + \frac{2\eta\rho_{\text{radiaton}}}{M_{Pl}^4}}-1\right)H_{\text{GR}}\,,
\end{equation}
giving
\begin{equation}
    \frac{\delta T_{\text{freeze}}}{T_{\text{freeze}}} = \frac{\left(\sqrt{1 + \frac{2\eta\rho_{\text{radiaton}}}{M_{Pl}^4}}-1\right)\left(\pi \sqrt{g_{\text{BBN}}}\right)}{15\sqrt{10}c_q T^3_{\text{freeze}}M_{Pl}}<4.7\times 10^{-4}\,, 
\end{equation}
leading to the constraint
\begin{equation}
    \eta<8.87644\times 10^{82} \xrightarrow{} \eta'<4.64484\times 10^{-28} \,.
\end{equation}

\subsection{Coupling between $\partial_\mu (\mathcal{T}^2)$ and $J^\mu$}\label{subsec:T2coupling}
{\it (ii) GR}:\label{sec:T2 GR}

We will first examine the impact of the interaction term \eqref{eq:T2 asymmetry} in GR. This step allows us to better understand the specific contributions arising from the scalar term $\mathcal{T}^2$. By establishing a clear baseline within the framework of GR, we can more effectively identify and interpret the deviations introduced by the extended theory. In this case, we can use Eqs. \eqref{eq:Friedmann n=1/2}, \eqref{eq:Acceleration n=1/2}. \eqref{eq:rho n=1/2 R} and \eqref{eq:rho_0 R n=1-2} with $\eta=0$ and the time derivative of Eq. \eqref{eq:T2} given by
\begin{equation}\label{eq:dot T^2 eta=0}
    \dot{\mathcal{T}}^2 = - \frac{16}{3}\left(\rho_{0}(\eta=0)\right)^2 t^{-5} \, ,
\end{equation}
to obtain the asymmetry (with $\epsilon=1$)
\begin{equation}\label{eq:asymmetry T2 n=1/2}
     \frac{n_b}{s} \simeq \frac{20\epsilon\pi^3 g_\ast^{3/2}g_b}{(30)^{5/2} M_{\ast}^8(\rho_{0}(0))^{1/2}} T^9_D \, ,
\end{equation}
where we found that for $T_D=1.48\times 10^{15}\ \text{GeV}$, $\epsilon=1$ and $M_{\ast} = 2\times 10^{16}\ \text{GeV}$ the model predicts successful baryogenesis with $n_b/s \approx 8.7904 \times 10^{-11}$, a notable result as no deviation from GR is needed to obtain a valid asymmetry. The scalar term $\mathcal{T}^2$, being proportional to $\rho^2 (1+3w^2)$, remains nonzero during the radiation-dominated epoch, unlike interaction terms such as Eq. \eqref{eq:GB asymmetry}, which vanish in this regime and cease contributing to baryon asymmetry. This persistence of $\mathcal{T}^2$ addresses a key limitation of conventional baryogenesis mechanisms.\\

{\it (ii) Model $n=1/2$}:\label{sec:T2 n 1/2}

Using now Eqs. \eqref{eq:Friedmann n=1/2}, \eqref{eq:Acceleration n=1/2}, \eqref{eq:rho n=1/2 R}, \eqref{eq:rho_0 R n=1-2} and \eqref{eq:dot T^2 eta=0} with $\eta\neq0$ leads to the asymmetry
\begin{equation}
     \frac{n_b}{s} \simeq \frac{20\epsilon\pi^3 g_\ast^{3/2}g_b}{(30)^{5/2} M_{\ast}^8(\rho_0(\eta))^{1/2}} T^9_D \, .
\end{equation}

As expected, this result is nearly identical to Eq. \eqref{eq:asymmetry T2 n=1/2}, with the key distinction that $ \rho_0 $ now depends on $ \eta $, thereby encapsulating the entire dependence of the asymmetry on $ \eta $. Taking into account the constraint from BBN, numerical computations showed that in the case of fixed $ M_\ast $ and $ T_D $, $ \eta $ exhibits a linear relationship with the baryon asymmetry where we found $n_b/s (\eta=0.001) \simeq 8.804 \times 10^{-11}$ for $T_D=7.984\times 10^{14}$ GeV, $\epsilon=1$ and $M_{\ast}=1\times 10^{16}$ GeV.

Considering now the constrained case $\eta_\ast$, the scalar $\mathcal{T}^2$ has the form
\begin{equation}
    \mathcal{T}^2(w) = (1+3w^2) \rho_c^2 t^{-8\lambda} \, ,
\end{equation}
and by using its time derivative and Eq. \eqref{eq:time 1-2 constraint} we arrive at the asymmetry
\begin{equation}
    \frac{n_b}{s} \simeq \epsilon\frac{g_b \lambda (1+3w^2) (\pi^2 g_\ast)^{1+{1}/{4\lambda}}} {(30)^{1+{1}/{4\lambda}}\rho_c^{{1}/4{\lambda}} M_\ast^8 } T_D^{7+\frac{1}{\lambda}} \, .
\end{equation}

For $ M_{\ast} = 1 \times 10^{16} $ GeV, $ T_D = 10^{14} $ GeV,  $\epsilon=1$ and $\rho_\text{c}=7.028\times 10 ^{36}$ we obtained successful baryon asymmetry $ n_b/s \simeq 8.8008 \times 10^{-11} $. Once more due to $\eta_\ast = -0.125$ this case is not compatible with BBN constraints.\\

{\it (iii) Model $n=1$}:\label{sec:T2 n 1}

For the $n=1$ case, using the trace of the field equations \eqref{eq:FE trace} the time derivative of $\mathcal{T}^2$ is
\begin{equation}\label{dot T^2 n=1}
    \dot{\mathcal{T}}^2 = \frac{8}{3} \dot{\rho} \rho \,.
\end{equation}

Due to the non-standard form of the continuity equation Eq.\eqref{eq:modified continuity}, it is expected that $\rho$ will deviate from the standard form $\rho \propto t^{-2}$. These deviations directly impact the interaction term \eqref{eq:T2 asymmetry} as it is proportional to $\rho^2$. Using Eq. \eqref{eq:Acc n=1} with \eqref{eq:H n=1} and considering the leading orders of $\rho$ one has
\begin{equation}
    \rho(t) \simeq \sqrt{\frac{3}{8\eta'}} t^{-1}\, ,
\end{equation}
leading to
\begin{equation}
    \dot{\mathcal{T}}^2 = -\eta'^{-1} t^{-3} \,,
\end{equation}
and to the decoupling time
\begin{equation}\label{decoupling time T^2 n=1}
    t_D = \sqrt{\frac{3}{8\eta'}} 30 \pi^{-2} g_\ast^{-1} T_D^{-4} \, .
\end{equation}

Combining all these results the asymmetry for $n=1$ is given by
\begin{equation}
     \frac{n_b}{s} \simeq \epsilon\frac{(8/3)^{3/2} g_b g_\ast^2 \pi^4 \eta^{1/2}}{7200 M_{Pl}^3 M_\ast^8} T_D^{11} \, .
\end{equation}

Analyzing the previous expression, we observe that for fixed $ M_\ast $ and $ T_D $, a higher $ \eta $ results in a greater asymmetry. Compared to other asymmetries, this one exhibits the most direct dependence on $ \eta $, making it the most sensitive to $ \mathcal{T}^2 $ contributions. For high decoupling temperatures, a low $\eta$, translating into a smaller influence from $\mathcal{T}^2$, facilitates a successful asymmetry. Conversely, at low $T_D$, a high $\eta$, which corresponds to enhanced contributions from $\mathcal{T}^2$, is required to achieve a positive outcome. Furthermore, from the action \eqref{eq:actionsf(R,T2)}, the coupling constant of $\mathcal{T}^2$, for $n=1$, is given by $\eta'=\eta/M_{Pl}^{4}$ and, compared to the GR coupling constant, for small $\eta$, the GR coupling remains significantly larger than $\eta'$. From the cosmological perspective, at high energies $\rho^2$ dominates and has an enhanced contribution when $\eta$ is sufficiently large to accommodate the suppressing factor of $M_{Pl}^6$. In this case, for $\epsilon=1$, $T_D = 1 \times 10^{12}$ GeV and $M_\ast = 1\times 10^{16}$ GeV we obtained the asymmetry $\frac{n_b}{s}(\eta=989.7 \times M_{Pl}^4) \simeq 8.8002 \times 10^{-11}$ a result that solves the asymmetry dilemma while being in agreement with BBN constraints. This result demonstrates the case where contributions from $\mathcal{T}^2$ are predominant in addition to effective.

\section{Discussion and conclusions}\label{sec:Summary and conclusion}

Before discussing the results, it is essential to clarify a key aspect: gravitational baryogenesis and its extensions require $B$ or $B-L$ violation. In this work, our primary focus is on the effects of $\mathcal{T}^2$, i.e, higher-order matter contributions to baryogenesis, while the specific $B$ or $B-L$ violation mechanisms will be examined in detail in a future study. Here, we provide only a general framework for their incorporation.

In an expanding Universe, baryon asymmetry can be generated in thermal equilibrium when the interaction rate $\Gamma$ exceeds the Hubble expansion rate $H$, i.e., $\Gamma \gg H$ for temperatures $T > T_D$. During this phase, $B$ or $B-L$ number-violating processes remain active. As the Universe cools to $T = T_D$, where $\Gamma \approx H$, these processes decouple. For $T < T_D$, when $H > \Gamma$, the baryon asymmetry becomes effectively conserved. In general, integrating a baryon ($B$) or baryon-lepton ($B-L$) violation mechanism requires establishing a relationship between temperature and the interaction rate $\Gamma$ of the $B$-violating process (or $B-L$). This connection links the parameters of the violation mechanism, such as the mass of a gauge boson mediating the interaction, to the gravitational baryogenesis parameter $T_D$ within a cosmological framework. Consider a given $B$/$B-l$ violating mechanism with an interaction rate 
\begin{equation}
    \Gamma_{B/B-L} = f(M_\chi,\ldots,\Lambda) g(T)\,,
\end{equation}
where $f(M_\chi, \ldots, \Lambda)$ encapsulates the dependence on the specific violation mechanism, incorporating parameters such as cutoff scales and the masses of mediating particles. The $g(T)$ depicts the dependence of the interaction rate on the temperature. By using $ \Gamma \approx H $ and the relevant cosmological equations, one can express the decoupling temperature $ T_D $ in terms of the parameters governing the $ B $/$ B-L $ violating interaction giving information about this mechanism and allowing to consider constraints from particle physics.

In this paper, we explored the implications for baryogenesis of high-order matter contributions in gravity  by incorporating power functions of the scalar $ \mathcal{T}^2 $ into GR. Within $ f(R,\mathcal{T}^2) $ gravity, this is achieved by the consideration of the class of models $ f(R,\mathcal{T}^2) = R + \eta'(\mathcal{T}^2)^n $ where we considered the models $n=\{\frac{1}{2},1\}$. We employed the gravitational baryogenesis mechanism and proposed an extension in which a novel interaction term is introduced, analogous to gravitational baryogenesis but incorporating $ \mathcal{T}^2 $ instead of the Ricci scalar. The cosmological equations for each model were derived and analyzed, serving as the fundamental building blocks for studying gravitational baryogenesis. We also explored a constrained case where the energy density was imposed to follow $\rho \propto a^{-4}$ given that we assume baryogenesis to occur during the radiation epoch. 

From the two models, $n=\frac{1}{2}$ was the one with the most successful cases and interesting results. Consisting in a simple deviation from GR, this model achieved successful results without requiring extreme contributions from $\mathcal{T}^2$ as only a small $\eta$ is necessary. Furthermore, this model exhibits the standard evolution of $\rho$ corresponding to radiation, as the modifications in the expansion rate are counterbalanced by adjustments in the continuity equation, resulting in their mutual cancellation. 

The $n=1$ model was only unable to obtain a positive outcome for the case of the coupling between the derivative of Ricci scalar and the baryonic current, presenting the same expansion rate as GR during the radiation epoch that leads to $\dot{R}=0$. This model has a quadratic energy density component in the cosmological equations, a feature that for baryogenesis revealed great interest being a fundamental aspect for the positive results obtained. Models with higher values of $n$ would lead the cosmological to have the form
\begin{equation}
H^2=B\frac{\eta' \rho ^{2n}}{3} \, ,
\end{equation}
\begin{equation}
\dot{H}+H^2=-C\frac{\eta'
\rho ^{2n}}{3} \, ,
\end{equation}
where $B$ and $C$ are constants, due to the fact that at high energies $\rho^{2n}$ dominates over $\rho$~\cite{Board:2017ign}. More complex models, which consider direct couplings between $R$ and $\mathcal{T}^2$, are also an interesting avenue to explore for future studies, as the non-minimal coupling between geometry and matter leads exotic effects such as gravitationally induced particle creation.

The interaction term \eqref{eq:T2 asymmetry} showed to have a crucial role in the context of baryogenesis, yielding successful results across all the scenarios, including GR and the modified gravity cases such as $n=1/2$ and $n=1$. Notably, this term effectively couples the baryon current $J^\mu_B$ with $\partial_\mu (\rho^2)$, highlighting a key mechanism through which high-energy regimes can significantly enhance baryon asymmetry. The quadratic dependence on $\rho$ underscores its potential to amplify contributions at early cosmological epochs, making it a compelling ingredient in scenarios of baryogenesis. Future studies exploring possible back-reactions of the interaction \eqref{eq:T2 asymmetry} and its consequences to gravitational waves are an interesting path to better understand this term and its consequences.

The positive results obtained in this study support the notion that, in high-energy regimes characteristic of the early Universe, higher-order matter contributions can play a crucial role in realizing successful baryogenesis. These findings highlight the importance of extending the standard gravitational framework to include nonlinear matter terms, particularly in scenarios beyond GR. In parallel, the favorable synergy arising from the cosmological modifications induced by $\mathcal{T}^2$ further strengthens the theoretical appeal of these extensions. The intrinsic nonlinearity of these models not only contributes to addressing fundamental questions regarding matter–antimatter asymmetry but also offers a promising avenue for understanding how gravity may behave under extreme conditions.

\section*{Acknowledgments}
The authors are grateful to the anonymous referee for valuable feedback. The authors acknowledge funding from the Funda\c{c}\~{a}o para a Ci\^{e}ncia e a Tecnologia (FCT) through the research grants UIDB/04434/2020, UIDP/04434/2020 and PTDC/FIS-AST/0054/2021.
FSNL also acknowledges support from the Funda\c{c}\~{a}o para a Ci\^{e}ncia e a Tecnologia (FCT) Scientific Employment Stimulus contract with reference CEECINST/00032/2018.

\bibliographystyle{elsarticle-num} 
\bibliography{biblio}

\begin{thebibliography}{10}
\expandafter\ifx\csname url\endcsname\relax
  \def\url#1{\texttt{#1}}\fi
\expandafter\ifx\csname urlprefix\endcsname\relax\def\urlprefix{URL }\fi
\expandafter\ifx\csname href\endcsname\relax
  \def\href#1#2{#2} \def\path#1{#1}\fi

\bibitem{c75ffd80-cea5-30a0-aee9-19091a4f0a9f}
F.~Wilczek, \href{http://www.jstor.org/stable/24966477}{The cosmic asymmetry
  between matter and antimatter}, Scientific American 243~(6) (1980) 82--91.
\newline\urlprefix\url{http://www.jstor.org/stable/24966477}

\bibitem{Burles:2000ju}
S.~Burles, K.~M. Nollett, M.~S. Turner, {What is the BBN prediction for the
  baryon density and how reliable is it?}, Phys. Rev. D 63 (2001) 063512.
\newblock \href {http://arxiv.org/abs/astro-ph/0008495}
  {\path{arXiv:astro-ph/0008495}}, \href
  {https://doi.org/10.1103/PhysRevD.63.063512}
  {\path{doi:10.1103/PhysRevD.63.063512}}.

\bibitem{WMAP:2003ivt}
C.~L. Bennett, et~al., {First year Wilkinson Microwave Anisotropy Probe (WMAP)
  observations: Preliminary maps and basic results}, Astrophys. J. Suppl. 148
  (2003) 1--27.
\newblock \href {http://arxiv.org/abs/astro-ph/0302207}
  {\path{arXiv:astro-ph/0302207}}, \href {https://doi.org/10.1086/377253}
  {\path{doi:10.1086/377253}}.

\bibitem{Burles:2000zk}
S.~Burles, K.~M. Nollett, M.~S. Turner, {Big bang nucleosynthesis predictions
  for precision cosmology}, Astrophys. J. Lett. 552 (2001) L1--L6.
\newblock \href {http://arxiv.org/abs/astro-ph/0010171}
  {\path{arXiv:astro-ph/0010171}}, \href {https://doi.org/10.1086/320251}
  {\path{doi:10.1086/320251}}.

\bibitem{WMAP:2003ogi}
C.~L. Bennett, et~al., {The Microwave Anisotropy Probe (MAP) mission},
  Astrophys. J. 583 (2003) 1--23.
\newblock \href {http://arxiv.org/abs/astro-ph/0301158}
  {\path{arXiv:astro-ph/0301158}}, \href {https://doi.org/10.1086/345346}
  {\path{doi:10.1086/345346}}.

\bibitem{Planck:2018vyg}
N.~Aghanim, et~al., {Planck 2018 results. VI. Cosmological parameters}, Astron.
  Astrophys. 641 (2020) A6, [Erratum: Astron.Astrophys. 652, C4 (2021)].
\newblock \href {http://arxiv.org/abs/1807.06209} {\path{arXiv:1807.06209}},
  \href {https://doi.org/10.1051/0004-6361/201833910}
  {\path{doi:10.1051/0004-6361/201833910}}.

\bibitem{Fields:2019pfx}
B.~D. Fields, K.~A. Olive, T.-H. Yeh, C.~Young, {Big-Bang Nucleosynthesis after
  Planck}, JCAP 03 (2020) 010, [Erratum: JCAP 11, E02 (2020)].
\newblock \href {http://arxiv.org/abs/1912.01132} {\path{arXiv:1912.01132}},
  \href {https://doi.org/10.1088/1475-7516/2020/03/010}
  {\path{doi:10.1088/1475-7516/2020/03/010}}.

\bibitem{ParticleDataGroup:2020ssz}
P.~A. Zyla, et~al., {Review of Particle Physics}, PTEP 2020~(8) (2020) 083C01.
\newblock \href {https://doi.org/10.1093/ptep/ptaa104}
  {\path{doi:10.1093/ptep/ptaa104}}.

\bibitem{Riotto:1999yt}
A.~Riotto, M.~Trodden, {Recent progress in baryogenesis}, Ann. Rev. Nucl. Part.
  Sci. 49 (1999) 35--75.
\newblock \href {http://arxiv.org/abs/hep-ph/9901362}
  {\path{arXiv:hep-ph/9901362}}, \href
  {https://doi.org/10.1146/annurev.nucl.49.1.35}
  {\path{doi:10.1146/annurev.nucl.49.1.35}}.

\bibitem{Shaposhnikov:2009zzb}
M.~Shaposhnikov, {Baryogenesis}, J. Phys. Conf. Ser. 171 (2009) 012005.
\newblock \href {https://doi.org/10.1088/1742-6596/171/1/012005}
  {\path{doi:10.1088/1742-6596/171/1/012005}}.

\bibitem{Morrissey:2012db}
D.~E. Morrissey, M.~J. Ramsey-Musolf, {Electroweak baryogenesis}, New J. Phys.
  14 (2012) 125003.
\newblock \href {http://arxiv.org/abs/1206.2942} {\path{arXiv:1206.2942}},
  \href {https://doi.org/10.1088/1367-2630/14/12/125003}
  {\path{doi:10.1088/1367-2630/14/12/125003}}.

\bibitem{Pereira:2023xiw}
D.~S. Pereira, J.~a. Ferraz, F.~S.~N. Lobo, J.~P. Mimoso, {Baryogenesis: A
  Symmetry Breaking in the Primordial Universe Revisited}, Symmetry 16~(1)
  (2024) 13.
\newblock \href {http://arxiv.org/abs/2312.14080} {\path{arXiv:2312.14080}},
  \href {https://doi.org/10.3390/sym16010013} {\path{doi:10.3390/sym16010013}}.

\bibitem{Davoudiasl:2004gf}
H.~Davoudiasl, R.~Kitano, G.~D. Kribs, H.~Murayama, P.~J. Steinhardt,
  {Gravitational baryogenesis}, Phys. Rev. Lett. 93 (2004) 201301.
\newblock \href {http://arxiv.org/abs/hep-ph/0403019}
  {\path{arXiv:hep-ph/0403019}}, \href
  {https://doi.org/10.1103/PhysRevLett.93.201301}
  {\path{doi:10.1103/PhysRevLett.93.201301}}.

\bibitem{Carroll:2005dj}
S.~M. Carroll, J.~Shu, {Models of baryogenesis via spontaneous Lorentz
  violation}, Phys. Rev. D 73 (2006) 103515.
\newblock \href {http://arxiv.org/abs/hep-ph/0510081}
  {\path{arXiv:hep-ph/0510081}}, \href
  {https://doi.org/10.1103/PhysRevD.73.103515}
  {\path{doi:10.1103/PhysRevD.73.103515}}.

\bibitem{Li:2006ss}
M.~Li, J.-Q. Xia, H.~Li, X.~Zhang, {Cosmological CPT violation,
  baryo/leptogenesis and CMB polarization}, Phys. Lett. B 651 (2007) 357--362.
\newblock \href {http://arxiv.org/abs/hep-ph/0611192}
  {\path{arXiv:hep-ph/0611192}}, \href
  {https://doi.org/10.1016/j.physletb.2007.06.050}
  {\path{doi:10.1016/j.physletb.2007.06.050}}.

\bibitem{Li:2008tma}
M.~Li, X.~Zhang, {Cosmological CPT violating effect on CMB polarization}, Phys.
  Rev. D 78 (2008) 103516.
\newblock \href {http://arxiv.org/abs/0810.0403} {\path{arXiv:0810.0403}},
  \href {https://doi.org/10.1103/PhysRevD.78.103516}
  {\path{doi:10.1103/PhysRevD.78.103516}}.

\bibitem{Xia:2008si}
J.-Q. Xia, H.~Li, G.-B. Zhao, X.~Zhang, {Testing CPT Symmetry with CMB
  Measurements: Update after WMAP5}, Astrophys. J. Lett. 679 (2008) L61--L63.
\newblock \href {http://arxiv.org/abs/0803.2350} {\path{arXiv:0803.2350}},
  \href {https://doi.org/10.1086/589447} {\path{doi:10.1086/589447}}.

\bibitem{Mavromatos:2013vqa}
N.~E. Mavromatos, {Violation of CPT Invariance in the Early Universe and
  Leptogenesis/Baryogenesis}, J. Phys. Conf. Ser. 447 (2013) 012016.
\newblock \href {https://doi.org/10.1088/1742-6596/447/1/012016}
  {\path{doi:10.1088/1742-6596/447/1/012016}}.

\bibitem{Mavromatos:2013boa}
N.~E. Mavromatos, S.~Sarkar, {CPT-violating leptogenesis induced by
  gravitational backgrounds}, J. Phys. Conf. Ser. 442 (2013) 012020.
\newblock \href {https://doi.org/10.1088/1742-6596/442/1/012020}
  {\path{doi:10.1088/1742-6596/442/1/012020}}.

\bibitem{McDonald:2014yfg}
J.~I. McDonald, G.~M. Shore, {Gravitational leptogenesis, C, CP and strong
  equivalence}, JHEP 02 (2015) 076.
\newblock \href {http://arxiv.org/abs/1411.3669} {\path{arXiv:1411.3669}},
  \href {https://doi.org/10.1007/JHEP02(2015)076}
  {\path{doi:10.1007/JHEP02(2015)076}}.

\bibitem{Mavromatos:2017gyn}
N.~E. Mavromatos, {Models and (some) Searches for CPT Violation: From Early
  Universe to the Present Era}, J. Phys. Conf. Ser. 873~(1) (2017) 012006.
\newblock \href {https://doi.org/10.1088/1742-6596/873/1/012006}
  {\path{doi:10.1088/1742-6596/873/1/012006}}.

\bibitem{Zhai:2020vob}
H.~Zhai, S.-Y. Li, M.~Li, H.~Li, X.~Zhang, {The effects on CMB power spectra
  and bispectra from the polarization rotation and its correlations with
  temperature and E-polarization}, JCAP 12 (2020) 051.
\newblock \href {http://arxiv.org/abs/2006.01811} {\path{arXiv:2006.01811}},
  \href {https://doi.org/10.1088/1475-7516/2020/12/051}
  {\path{doi:10.1088/1475-7516/2020/12/051}}.

\bibitem{Li:2004hh}
H.~Li, M.-z. Li, X.-m. Zhang, {Gravitational leptogenesis and neutrino mass
  limit}, Phys. Rev. D 70 (2004) 047302.
\newblock \href {http://arxiv.org/abs/hep-ph/0403281}
  {\path{arXiv:hep-ph/0403281}}, \href
  {https://doi.org/10.1103/PhysRevD.70.047302}
  {\path{doi:10.1103/PhysRevD.70.047302}}.

\bibitem{Lambiase:2006dq}
G.~Lambiase, G.~Scarpetta, {Baryogenesis in f(R): Theories of Gravity}, Phys.
  Rev. D 74 (2006) 087504.
\newblock \href {http://arxiv.org/abs/astro-ph/0610367}
  {\path{arXiv:astro-ph/0610367}}, \href
  {https://doi.org/10.1103/PhysRevD.74.087504}
  {\path{doi:10.1103/PhysRevD.74.087504}}.

\bibitem{Lambiase:2012tn}
G.~Lambiase, S.~Mohanty, L.~Pizza, {Consequences of f(R)-theories of gravity on
  gravitational leptogenesis}, Gen. Rel. Grav. 45 (2013) 1771--1785.
\newblock \href {http://arxiv.org/abs/1212.6026} {\path{arXiv:1212.6026}},
  \href {https://doi.org/10.1007/s10714-013-1555-4}
  {\path{doi:10.1007/s10714-013-1555-4}}.

\bibitem{MohseniSadjadi:2007qk}
H.~Mohseni~Sadjadi, {Multicomponent solution in modified theory of gravity},
  Phys. Rev. D 77 (2008) 103501.
\newblock \href {http://arxiv.org/abs/0710.3308} {\path{arXiv:0710.3308}},
  \href {https://doi.org/10.1103/PhysRevD.77.103501}
  {\path{doi:10.1103/PhysRevD.77.103501}}.

\bibitem{Bhattacharjee:2020jfk}
S.~Bhattacharjee, {Gravitational baryogenesis in extended teleparallel theories
  of gravity}, Phys. Dark Univ. 30 (2020) 100612.
\newblock \href {http://arxiv.org/abs/2005.05534} {\path{arXiv:2005.05534}},
  \href {https://doi.org/10.1016/j.dark.2020.100612}
  {\path{doi:10.1016/j.dark.2020.100612}}.

\bibitem{Mojahed:2024yus}
M.~A. Mojahed, K.~Schmitz, X.-J. Xu, {Gravitational chargegenesis} (9 2024).
\newblock \href {http://arxiv.org/abs/2409.10605} {\path{arXiv:2409.10605}}.

\bibitem{Jaybhaye:2023lgr}
L.~V. Jaybhaye, S.~Bhattacharjee, P.~K. Sahoo, {Baryogenesis in f(R,Lm)
  gravity}, Phys. Dark Univ. 40 (2023) 101223.
\newblock \href {http://arxiv.org/abs/2304.02482} {\path{arXiv:2304.02482}},
  \href {https://doi.org/10.1016/j.dark.2023.101223}
  {\path{doi:10.1016/j.dark.2023.101223}}.

\bibitem{Baffou:2018hpe}
E.~H. Baffou, M.~J.~S. Houndjo, D.~A. Kanfon, I.~G. Salako, {$f(R,T)$ models
  applied to baryogenesis}, Eur. Phys. J. C 79~(2) (2019) 112.
\newblock \href {http://arxiv.org/abs/1808.01917} {\path{arXiv:1808.01917}},
  \href {https://doi.org/10.1140/epjc/s10052-019-6559-0}
  {\path{doi:10.1140/epjc/s10052-019-6559-0}}.

\bibitem{Nozari:2018ift}
K.~Nozari, F.~Rajabi, {Baryogenesis in $f(R,T)$ Gravity}, Commun. Theor. Phys.
  70~(4) (2018) 451.
\newblock \href {https://doi.org/10.1088/0253-6102/70/4/451}
  {\path{doi:10.1088/0253-6102/70/4/451}}.

\bibitem{Sahoo:2019pat}
P.~K. Sahoo, S.~Bhattacharjee, {Gravitational Baryogenesis in Non-Minimal
  Coupled $f(R,T)$ Gravity}, Int. J. Theor. Phys. 59~(5) (2020) 1451--1459.
\newblock \href {http://arxiv.org/abs/1907.13460} {\path{arXiv:1907.13460}},
  \href {https://doi.org/10.1007/s10773-020-04414-3}
  {\path{doi:10.1007/s10773-020-04414-3}}.

\bibitem{Odintsov:2016hgc}
S.~D. Odintsov, V.~K. Oikonomou, {Gauss\textendash{}Bonnet gravitational
  baryogenesis}, Phys. Lett. B 760 (2016) 259--262.
\newblock \href {http://arxiv.org/abs/1607.00545} {\path{arXiv:1607.00545}},
  \href {https://doi.org/10.1016/j.physletb.2016.06.074}
  {\path{doi:10.1016/j.physletb.2016.06.074}}.

\bibitem{Arbuzova:2023rri}
E.~Arbuzova, A.~Dolgov, K.~Dutta, R.~Rangarajan, {Gravitational Baryogenesis:
  Problems and Possible Resolution}, Symmetry 15~(2) (2023) 404.
\newblock \href {http://arxiv.org/abs/2301.08322} {\path{arXiv:2301.08322}},
  \href {https://doi.org/10.3390/sym15020404} {\path{doi:10.3390/sym15020404}}.

\bibitem{Nojiri:2010wj}
S.~Nojiri, S.~D. Odintsov, {Unified cosmic history in modified gravity: from
  F(R) theory to Lorentz non-invariant models}, Phys. Rept. 505 (2011) 59--144.
\newblock \href {http://arxiv.org/abs/1011.0544} {\path{arXiv:1011.0544}},
  \href {https://doi.org/10.1016/j.physrep.2011.04.001}
  {\path{doi:10.1016/j.physrep.2011.04.001}}.

\bibitem{Nojiri:2017ncd}
S.~Nojiri, S.~D. Odintsov, V.~K. Oikonomou, {Modified Gravity Theories on a
  Nutshell: Inflation, Bounce and Late-time Evolution}, Phys. Rept. 692 (2017)
  1--104.
\newblock \href {http://arxiv.org/abs/1705.11098} {\path{arXiv:1705.11098}},
  \href {https://doi.org/10.1016/j.physrep.2017.06.001}
  {\path{doi:10.1016/j.physrep.2017.06.001}}.

\bibitem{CANTATA:2021asi}
Y.~Akrami, et~al., {Modified Gravity and Cosmology. An Update by the CANTATA
  Network}, Springer, 2021.
\newblock \href {http://arxiv.org/abs/2105.12582} {\path{arXiv:2105.12582}},
  \href {https://doi.org/10.1007/978-3-030-83715-0}
  {\path{doi:10.1007/978-3-030-83715-0}}.

\bibitem{Planck:2018jri}
Y.~Akrami, et~al., {Planck 2018 results. X. Constraints on inflation}, Astron.
  Astrophys. 641 (2020) A10.
\newblock \href {http://arxiv.org/abs/1807.06211} {\path{arXiv:1807.06211}},
  \href {https://doi.org/10.1051/0004-6361/201833887}
  {\path{doi:10.1051/0004-6361/201833887}}.

\bibitem{Dzhunushaliev:2013nea}
V.~Dzhunushaliev, V.~Folomeev, B.~Kleihaus, J.~Kunz, {Modified gravity from the
  quantum part of the metric}, Eur. Phys. J. C 74 (2014) 2743.
\newblock \href {http://arxiv.org/abs/1312.0225} {\path{arXiv:1312.0225}},
  \href {https://doi.org/10.1140/epjc/s10052-014-2743-4}
  {\path{doi:10.1140/epjc/s10052-014-2743-4}}.

\bibitem{Maartens:2010ar}
R.~Maartens, K.~Koyama, {Brane-World Gravity}, Living Rev. Rel. 13 (2010) 5.
\newblock \href {http://arxiv.org/abs/1004.3962} {\path{arXiv:1004.3962}},
  \href {https://doi.org/10.12942/lrr-2010-5} {\path{doi:10.12942/lrr-2010-5}}.

\bibitem{Hack:2012qf}
T.-P. Hack, V.~Moretti, {On the Stress-Energy Tensor of Quantum Fields in
  Curved Spacetimes - Comparison of Different Regularization Schemes and
  Symmetry of the Hadamard/Seeley-DeWitt Coefficients}, J. Phys. A 45 (2012)
  374019.
\newblock \href {http://arxiv.org/abs/1202.5107} {\path{arXiv:1202.5107}},
  \href {https://doi.org/10.1088/1751-8113/45/37/374019}
  {\path{doi:10.1088/1751-8113/45/37/374019}}.

\bibitem{Ford:1997hb}
L.~H. Ford, {Quantum field theory in curved space-time}, in: {9th Jorge Andre
  Swieca Summer School: Particles and Fields}, 1997, pp. 345--388.
\newblock \href {http://arxiv.org/abs/gr-qc/9707062}
  {\path{arXiv:gr-qc/9707062}}.

\bibitem{Katirci:2013okf}
N.~Kat\i{}rc\i{}, M.~Kavuk, {$ f(R,T_{\mu\nu}T^{\mu\nu})$ gravity and
  Cardassian-like expansion as one of its consequences}, Eur. Phys. J. Plus 129
  (2014) 163.
\newblock \href {http://arxiv.org/abs/1302.4300} {\path{arXiv:1302.4300}},
  \href {https://doi.org/10.1140/epjp/i2014-14163-6}
  {\path{doi:10.1140/epjp/i2014-14163-6}}.

\bibitem{Board:2017ign}
C.~V.~R. Board, J.~D. Barrow, {Cosmological Models in Energy-Momentum-Squared
  Gravity}, Phys. Rev. D 96~(12) (2017) 123517, [Erratum: Phys.Rev.D 98, 129902
  (2018)].
\newblock \href {http://arxiv.org/abs/1709.09501} {\path{arXiv:1709.09501}},
  \href {https://doi.org/10.1103/PhysRevD.96.123517}
  {\path{doi:10.1103/PhysRevD.96.123517}}.

\bibitem{Roshan:2016mbt}
M.~Roshan, F.~Shojai, {Energy-Momentum Squared Gravity}, Phys. Rev. D 94~(4)
  (2016) 044002.
\newblock \href {http://arxiv.org/abs/1607.06049} {\path{arXiv:1607.06049}},
  \href {https://doi.org/10.1103/PhysRevD.94.044002}
  {\path{doi:10.1103/PhysRevD.94.044002}}.

\bibitem{Akarsu:2022abd}
O.~Akarsu, A.~K. Camlibel, N.~Katirci, I.~Semiz, N.~M. Uzun, {Weak field and
  slow motion limits in energy\textendash{}momentum powered gravity}, Phys.
  Dark Univ. 42 (2023) 101305.
\newblock \href {http://arxiv.org/abs/2210.04668} {\path{arXiv:2210.04668}},
  \href {https://doi.org/10.1016/j.dark.2023.101305}
  {\path{doi:10.1016/j.dark.2023.101305}}.

\bibitem{Akarsu:2023agp}
O.~Akarsu, M.~Bouhmadi-L\'opez, N.~Katirci, N.~M. Uzun, {Quadratic
  energy\textendash{}momentum squared gravity: Constraints from big bang
  nucleosynthesis}, Phys. Dark Univ. 45 (2024) 101505.
\newblock \href {http://arxiv.org/abs/2312.11453} {\path{arXiv:2312.11453}},
  \href {https://doi.org/10.1016/j.dark.2024.101505}
  {\path{doi:10.1016/j.dark.2024.101505}}.

\bibitem{Nazari:2022xhv}
E.~Nazari, M.~Roshan, I.~De~Martino, {Constraining energy-momentum-squared
  gravity by binary pulsar observations}, Phys. Rev. D 105~(4) (2022) 044014.
\newblock \href {http://arxiv.org/abs/2201.08578} {\path{arXiv:2201.08578}},
  \href {https://doi.org/10.1103/PhysRevD.105.044014}
  {\path{doi:10.1103/PhysRevD.105.044014}}.

\bibitem{Akarsu:2018zxl}
O.~Akarsu, J.~D. Barrow, S.~\c{C}\i{}k\i{}nto\u{g}lu, K.~Y. Ek\c{s}i,
  N.~Kat\i{}rc\i{}, {Constraint on energy-momentum squared gravity from neutron
  stars and its cosmological implications}, Phys. Rev. D 97~(12) (2018) 124017.
\newblock \href {http://arxiv.org/abs/1802.02093} {\path{arXiv:1802.02093}},
  \href {https://doi.org/10.1103/PhysRevD.97.124017}
  {\path{doi:10.1103/PhysRevD.97.124017}}.

\bibitem{HosseiniMansoori:2023zop}
S.~A. Hosseini~Mansoori, F.~Felegary, M.~Roshan, O.~Akarsu, M.~Sami, {T2-
  inflation: Sourced by energy\textendash{}momentum squared gravity}, Phys.
  Dark Univ. 42 (2023) 101360.
\newblock \href {http://arxiv.org/abs/2306.09181} {\path{arXiv:2306.09181}},
  \href {https://doi.org/10.1016/j.dark.2023.101360}
  {\path{doi:10.1016/j.dark.2023.101360}}.

\bibitem{HosseiniMansoori:2023mqh}
S.~A. Hosseini~Mansoori, F.~Felegray, A.~Talebian, M.~Sami, {PBHs and GWs from
  \ensuremath{\mathbb{T}}$^{2}$-inflation and NANOGrav 15-year data}, JCAP 08
  (2023) 067.
\newblock \href {http://arxiv.org/abs/2307.06757} {\path{arXiv:2307.06757}},
  \href {https://doi.org/10.1088/1475-7516/2023/08/067}
  {\path{doi:10.1088/1475-7516/2023/08/067}}.

\bibitem{Jang:2024jso}
D.~Jang, M.~R. Gangopadhyay, M.-K. Cheoun, T.~Kajino, M.~Sami, {Big Bang
  Nucleosynthesis constraints on the Energy-Momentum Squared Gravity: The
  $\mathbb{T}^{2}$ model} (2 2024).
\newblock \href {http://arxiv.org/abs/2402.01210} {\path{arXiv:2402.01210}}.

\bibitem{Barbar:2019rfn}
A.~H. Barbar, A.~M. Awad, M.~T. AlFiky, {Viability of bouncing cosmology in
  energy-momentum-squared gravity}, Phys. Rev. D 101~(4) (2020) 044058.
\newblock \href {http://arxiv.org/abs/1911.00556} {\path{arXiv:1911.00556}},
  \href {https://doi.org/10.1103/PhysRevD.101.044058}
  {\path{doi:10.1103/PhysRevD.101.044058}}.

\bibitem{Faraji:2021laz}
M.~Faraji, N.~Rashidi, K.~Nozari, {Inflation in energy-momentum squared gravity
  in light of Planck2018}, Eur. Phys. J. Plus 137~(5) (2022) 593.
\newblock \href {http://arxiv.org/abs/2107.13547} {\path{arXiv:2107.13547}},
  \href {https://doi.org/10.1140/epjp/s13360-022-02820-6}
  {\path{doi:10.1140/epjp/s13360-022-02820-6}}.

\bibitem{Nazari:2020gnu}
E.~Nazari, F.~Sarvi, M.~Roshan, {Generalized Energy-Momentum-Squared Gravity in
  the Palatini Formalism}, Phys. Rev. D 102~(6) (2020) 064016.
\newblock \href {http://arxiv.org/abs/2008.06681} {\path{arXiv:2008.06681}},
  \href {https://doi.org/10.1103/PhysRevD.102.064016}
  {\path{doi:10.1103/PhysRevD.102.064016}}.

\bibitem{Harko:2014pqa}
T.~Harko, {Thermodynamic interpretation of the generalized gravity models with
  geometry - matter coupling}, Phys. Rev. D 90~(4) (2014) 044067.
\newblock \href {http://arxiv.org/abs/1408.3465} {\path{arXiv:1408.3465}},
  \href {https://doi.org/10.1103/PhysRevD.90.044067}
  {\path{doi:10.1103/PhysRevD.90.044067}}.

\bibitem{Harko:2015pma}
T.~Harko, F.~S.~N. Lobo, J.~P. Mimoso, D.~Pav\'on, {Gravitational induced
  particle production through a nonminimal curvature\textendash{}matter
  coupling}, Eur. Phys. J. C 75 (2015) 386.
\newblock \href {http://arxiv.org/abs/1508.02511} {\path{arXiv:1508.02511}},
  \href {https://doi.org/10.1140/epjc/s10052-015-3620-5}
  {\path{doi:10.1140/epjc/s10052-015-3620-5}}.

\bibitem{Pinto:2022tlu}
M.~A.~S. Pinto, T.~Harko, F.~S.~N. Lobo, {Gravitationally induced particle
  production in scalar-tensor f(R,T) gravity}, Phys. Rev. D 106~(4) (2022)
  044043.
\newblock \href {http://arxiv.org/abs/2205.12545} {\path{arXiv:2205.12545}},
  \href {https://doi.org/10.1103/PhysRevD.106.044043}
  {\path{doi:10.1103/PhysRevD.106.044043}}.

\bibitem{Pinto:2023phl}
M.~A.~S. Pinto, T.~Harko, F.~S.~N. Lobo, {Irreversible Geometrothermodynamics
  of Open Systems in Modified Gravity}, Entropy 25~(6) (2023) 944.
\newblock \href {http://arxiv.org/abs/2306.13912} {\path{arXiv:2306.13912}},
  \href {https://doi.org/10.3390/e25060944} {\path{doi:10.3390/e25060944}}.

\bibitem{Cipriano:2023yhv}
R.~A.~C. Cipriano, T.~Harko, F.~S.~N. Lobo, M.~A.~S. Pinto, J.~a.~L. Rosa,
  {Gravitationally induced matter creation in scalar\textendash{}tensor
  f(R,T\ensuremath{\mu}\ensuremath{\nu}T\ensuremath{\mu}\ensuremath{\nu})
  gravity}, Phys. Dark Univ. 44 (2024) 101463.
\newblock \href {http://arxiv.org/abs/2310.15018} {\path{arXiv:2310.15018}},
  \href {https://doi.org/10.1016/j.dark.2024.101463}
  {\path{doi:10.1016/j.dark.2024.101463}}.

\bibitem{Kolb:1990vq}
E.~W. Kolb, {The Early Universe}, Vol.~69, Taylor and Francis, 2019.
\newblock \href {https://doi.org/10.1201/9780429492860}
  {\path{doi:10.1201/9780429492860}}.

\bibitem{Green:1987sp}
M.~B. Green, J.~H. Schwarz, E.~Witten, {SUPERSTRING THEORY. VOL. 1:
  INTRODUCTION}, Cambridge Monographs on Mathematical Physics, 1988.

\bibitem{Green:1987mn}
M.~B. Green, J.~H. Schwarz, E.~Witten, {SUPERSTRING THEORY. VOL. 2: LOOP
  AMPLITUDES, ANOMALIES AND PHENOMENOLOGY}, 1988.

\bibitem{Shtanov:2002mb}
Y.~Shtanov, V.~Sahni, {Bouncing brane worlds}, Phys. Lett. B 557 (2003) 1--6.
\newblock \href {http://arxiv.org/abs/gr-qc/0208047}
  {\path{arXiv:gr-qc/0208047}}, \href
  {https://doi.org/10.1016/S0370-2693(03)00179-5}
  {\path{doi:10.1016/S0370-2693(03)00179-5}}.

\bibitem{Ashtekar:2006uz}
A.~Ashtekar, T.~Pawlowski, P.~Singh, {Quantum Nature of the Big Bang: An
  Analytical and Numerical Investigation. I.}, Phys. Rev. D 73 (2006) 124038.
\newblock \href {http://arxiv.org/abs/gr-qc/0604013}
  {\path{arXiv:gr-qc/0604013}}, \href
  {https://doi.org/10.1103/PhysRevD.73.124038}
  {\path{doi:10.1103/PhysRevD.73.124038}}.

\bibitem{Bertolami:2007gv}
O.~Bertolami, C.~G. Boehmer, T.~Harko, F.~S.~N. Lobo, {Extra force in f(R)
  modified theories of gravity}, Phys. Rev. D 75 (2007) 104016.
\newblock \href {http://arxiv.org/abs/0704.1733} {\path{arXiv:0704.1733}},
  \href {https://doi.org/10.1103/PhysRevD.75.104016}
  {\path{doi:10.1103/PhysRevD.75.104016}}.

\bibitem{Schutz:1970my}
B.~F. Schutz, {Perfect Fluids in General Relativity: Velocity Potentials and a
  Variational Principle}, Phys. Rev. D 2 (1970) 2762--2773.
\newblock \href {https://doi.org/10.1103/PhysRevD.2.2762}
  {\path{doi:10.1103/PhysRevD.2.2762}}.

\bibitem{Bertolami:2008ab}
O.~Bertolami, F.~S.~N. Lobo, J.~Paramos, {Non-minimum coupling of perfect
  fluids to curvature}, Phys. Rev. D 78 (2008) 064036.
\newblock \href {http://arxiv.org/abs/0806.4434} {\path{arXiv:0806.4434}},
  \href {https://doi.org/10.1103/PhysRevD.78.064036}
  {\path{doi:10.1103/PhysRevD.78.064036}}.

\bibitem{Akarsu:2023lre}
O.~Akarsu, M.~Bouhmadi-L\'opez, N.~Kat\i{}rc\i{}, E.~Nazari, M.~Roshan, N.~M.
  Uzun, {Equivalence of matter-type modified gravity theories to general
  relativity with nonminimal matter interaction}, Phys. Rev. D 109~(10) (2024)
  104055.
\newblock \href {http://arxiv.org/abs/2306.11717} {\path{arXiv:2306.11717}},
  \href {https://doi.org/10.1103/PhysRevD.109.104055}
  {\path{doi:10.1103/PhysRevD.109.104055}}.

\bibitem{Pereira:2024ddu}
D.~S. Pereira, {Scalar-tensor Baryogenesis} (12 2024).
\newblock \href {http://arxiv.org/abs/2412.06984} {\path{arXiv:2412.06984}}.

\bibitem{Ahmad:2019jbm}
S.~Ahmad, A.~De~Felice, N.~Jaman, S.~Kuroyanagi, M.~Sami, {Baryogenesis in the
  paradigm of quintessential inflation}, Phys. Rev. D 100~(10) (2019) 103525.
\newblock \href {http://arxiv.org/abs/1908.03742} {\path{arXiv:1908.03742}},
  \href {https://doi.org/10.1103/PhysRevD.100.103525}
  {\path{doi:10.1103/PhysRevD.100.103525}}.

\bibitem{Rangarajan:2001yu}
R.~Rangarajan, D.~V. Nanopoulos, {Inflationary baryogenesis}, Phys. Rev. D 64
  (2001) 063511.
\newblock \href {http://arxiv.org/abs/hep-ph/0103348}
  {\path{arXiv:hep-ph/0103348}}, \href
  {https://doi.org/10.1103/PhysRevD.64.063511}
  {\path{doi:10.1103/PhysRevD.64.063511}}.

\bibitem{Cado:2023zbm}
Y.~Cado, C.~Englert, T.~Modak, M.~Quir\'os, {Baryogenesis in R2-Higgs
  inflation: The gravitational connection}, Phys. Rev. D 109~(4) (2024) 043026.
\newblock \href {http://arxiv.org/abs/2312.10414} {\path{arXiv:2312.10414}},
  \href {https://doi.org/10.1103/PhysRevD.109.043026}
  {\path{doi:10.1103/PhysRevD.109.043026}}.

\bibitem{Gunji:2023xme}
Y.~Gunji, K.~Ishiwata, T.~Yoshida, {Supersymmetric baryogenesis in a hybrid
  inflation model}, JHEP 08 (2023) 201.
\newblock \href {http://arxiv.org/abs/2303.05663} {\path{arXiv:2303.05663}},
  \href {https://doi.org/10.1007/JHEP08(2023)201}
  {\path{doi:10.1007/JHEP08(2023)201}}.

\bibitem{Gouttenoire:2021jhk}
Y.~Gouttenoire, G.~Servant, P.~Simakachorn, {Kination cosmology from scalar
  fields and gravitational-wave signatures} (11 2021).
\newblock \href {http://arxiv.org/abs/2111.01150} {\path{arXiv:2111.01150}}.

\bibitem{Yeh:2022heq}
T.-H. Yeh, J.~Shelton, K.~A. Olive, B.~D. Fields, {Probing physics beyond the
  standard model: limits from BBN and the CMB independently and combined}, JCAP
  10 (2022) 046.
\newblock \href {http://arxiv.org/abs/2207.13133} {\path{arXiv:2207.13133}},
  \href {https://doi.org/10.1088/1475-7516/2022/10/046}
  {\path{doi:10.1088/1475-7516/2022/10/046}}.

\bibitem{Cyburt:2015mya}
R.~H. Cyburt, B.~D. Fields, K.~A. Olive, T.-H. Yeh, {Big Bang Nucleosynthesis:
  2015}, Rev. Mod. Phys. 88 (2016) 015004.
\newblock \href {http://arxiv.org/abs/1505.01076} {\path{arXiv:1505.01076}},
  \href {https://doi.org/10.1103/RevModPhys.88.015004}
  {\path{doi:10.1103/RevModPhys.88.015004}}.

\bibitem{Park:2022lgf}
J.~Park, C.-m. Yun, M.-K. Cheoun, D.~Jang, {Oscillating cosmic evolution and
  constraints on big bang nucleosynthesis in the extended Starobinsky model},
  JCAP 05 (2023) 016.
\newblock \href {http://arxiv.org/abs/2212.11487} {\path{arXiv:2212.11487}},
  \href {https://doi.org/10.1088/1475-7516/2023/05/016}
  {\path{doi:10.1088/1475-7516/2023/05/016}}.

\bibitem{Mukhanov:2003xs}
V.~F. Mukhanov, {Nucleosynthesis without a computer}, Int. J. Theor. Phys. 43
  (2004) 669--693.
\newblock \href {http://arxiv.org/abs/astro-ph/0303073}
  {\path{arXiv:astro-ph/0303073}}, \href
  {https://doi.org/10.1023/B:IJTP.0000048169.69609.77}
  {\path{doi:10.1023/B:IJTP.0000048169.69609.77}}.

\bibitem{Capozziello:2017bxm}
S.~Capozziello, G.~Lambiase, E.~N. Saridakis, {Constraining f(T) teleparallel
  gravity by Big Bang Nucleosynthesis}, Eur. Phys. J. C 77~(9) (2017) 576.
\newblock \href {http://arxiv.org/abs/1702.07952} {\path{arXiv:1702.07952}},
  \href {https://doi.org/10.1140/epjc/s10052-017-5143-8}
  {\path{doi:10.1140/epjc/s10052-017-5143-8}}.

\bibitem{Asimakis:2021yct}
P.~Asimakis, S.~Basilakos, N.~E. Mavromatos, E.~N. Saridakis, {Big bang
  nucleosynthesis constraints on higher-order modified gravities}, Phys. Rev. D
  105~(8) (2022) 084010.
\newblock \href {http://arxiv.org/abs/2112.10863} {\path{arXiv:2112.10863}},
  \href {https://doi.org/10.1103/PhysRevD.105.084010}
  {\path{doi:10.1103/PhysRevD.105.084010}}.

\bibitem{Ghoshal:2021ief}
A.~Ghoshal, G.~Lambiase, {Constraints on Tsallis Cosmology from Big Bang
  Nucleosynthesis and Dark Matter Freeze-out} (4 2021).
\newblock \href {http://arxiv.org/abs/2104.11296} {\path{arXiv:2104.11296}}.

\bibitem{Coc:2003ce}
A.~Coc, E.~Vangioni-Flam, P.~Descouvemont, A.~Adahchour, C.~Angulo, {Updated
  Big Bang nucleosynthesis confronted to WMAP observations and to the abundance
  of light elements}, Astrophys. J. 600 (2004) 544--552.
\newblock \href {http://arxiv.org/abs/astro-ph/0309480}
  {\path{arXiv:astro-ph/0309480}}, \href {https://doi.org/10.1086/380121}
  {\path{doi:10.1086/380121}}.

\bibitem{Izotov:2003xn}
Y.~I. Izotov, T.~X. Thuan, {Systematic effects and a new determination of the
  primordial abundance of He-4 and dY/dZ from observations of blue compact
  galaxies}, Astrophys. J. 602 (2004) 200--230.
\newblock \href {http://arxiv.org/abs/astro-ph/0310421}
  {\path{arXiv:astro-ph/0310421}}, \href {https://doi.org/10.1086/380830}
  {\path{doi:10.1086/380830}}.

\end{thebibliography}

\end{document}